\documentclass[format=acmsmall]{acmart}

\usepackage[show]{chato-notes}

\usepackage{graphics}
\usepackage{amsfonts}
\usepackage{subfigure}
\usepackage{multirow}
\usepackage[algo2e,ruled,vlined]{algorithm2e}
\usepackage{graphics}
\usepackage{soul}
\usepackage{adjustbox}
\usepackage{enumitem}

\DeclareMathOperator*{\argmin}{arg\,min}

\newcommand{\db}      {\ensuremath{\mathcal{D}}\xspace}
\newcommand{\cache}   {\ensuremath{\mathcal{C}}\xspace}
\newcommand{\mindex}   {\ensuremath{\mathcal{M}}\xspace}
\newcommand{\results}   {\ensuremath{\mathcal{R}}\xspace}

\newcommand{\cast}{\textsc{CAsT}\xspace}

\newcommand{\msmarco}{\textsc{MS-MARCO}\xspace}

\newcommand{\cachealg}{\textsf{CACHE}\xspace}

\newcommand{\raf}[1]{\textcolor{black}{#1}}
\newcommand{\cris}[1]{\textcolor{black}{#1}}

\newcommand{\ophir}[1]{\textcolor{black}{#1}}
\newcommand{\fm}[1]{\textcolor{black}{#1}}

\newcommand{\acov}{\textsf{cov}}

\acmJournal{TWEB}

\begin{document}

\title{Caching Historical Embeddings in Conversational Search}

\author{Ophir Frieder}
\affiliation{%
  \institution{Georgetown University}}
\email{ophir@ir.cs.georgetown.edu}

\author{Ida Mele}
\affiliation{%
\institution{IASI-CNR}}
\email{ida.mele@iasi.cnr.it}

\author{Cristina Ioana Muntean}
\affiliation{%
\institution{ISTI-CNR}}
\email{cristina.muntean@isti.cnr.it}

\author{Franco Maria Nardini}
\affiliation{%
\institution{ISTI-CNR}}
\email{francomaria.nardini@isti.cnr.it}

\author{Raffaele Perego}
\affiliation{%
\institution{ISTI-CNR}}
\email{raffaele.perego@isti.cnr.it}

\author{Nicola Tonellotto}
\affiliation{%
  \institution{University of Pisa}}
  \email{nicola.tonellotto@unipi.it}


\begin{abstract}
Rapid response, namely low latency, is fundamental in search applications; it is particularly so in interactive search sessions, such as those encountered in conversational settings. An observation with a potential to reduce latency asserts that conversational queries exhibit a temporal locality in the lists of documents retrieved.  Motivated by this observation, we propose and evaluate a client-side document embedding cache, improving the responsiveness of conversational search systems. 
By leveraging state-of-the-art dense retrieval models to abstract document and query semantics, we cache the embeddings of documents retrieved for a topic introduced in the conversation, as they are likely relevant to successive queries. Our document embedding cache implements an efficient metric index, answering nearest-neighbor similarity queries by estimating the approximate result sets returned.  We demonstrate the efficiency achieved using our cache via reproducible experiments based on TREC CAsT datasets, achieving a hit rate of up to 75\% 
without degrading answer quality. Our achieved high cache hit rates significantly improve the responsiveness of conversational systems while likewise reducing the number of queries managed on the search back-end. 
\end{abstract}
\maketitle
\renewcommand{\shortauthors}{FILL}

\section{Introduction}\label{sec:introduction}


Conversational agents, fueled by language understanding advancements enabled by large contextualized language models, are drawing considerable attention~\cite{anand2020conversational,zamani2022conversational}. Multi-turn conversations commence with a main topic and evolve with differing facets of the initial topic or an abrupt shift to a new focus, possibly suggested by the content of the answers returned~\cite{Mele2021AdaptiveUR,DaltonEtAl2020}. 

\begin{figure}[htb]
\centering
\includegraphics[width=0.6\linewidth]{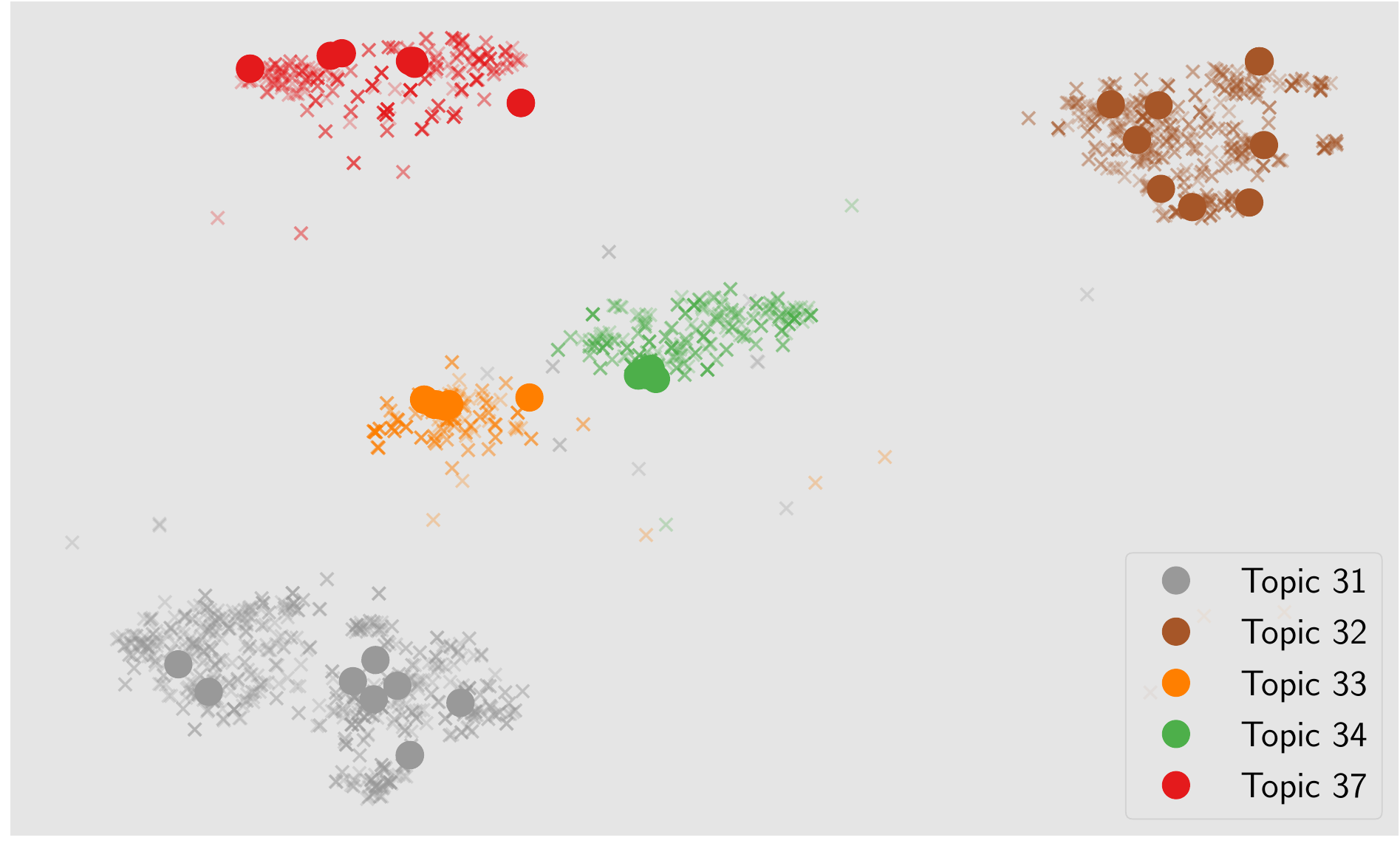}
\caption{2D visualization of  conversational queries ($\bullet$) and  corresponding relevant documents ($\times$) for 5 \cast 2019 topics.}\label{fig:motivation}
\end{figure}

A user drives such an interactive information-discovery process by submitting a query about a topic followed by a sequence of more specific queries, possibly aimed at clarifying some aspects of the topic. Documents relevant to the first query 
 are often relevant and helpful in answering subsequent 
queries.
This suggests the presence of temporal locality in the lists of results retrieved by conversational systems for successive queries issued by the same user in the same conversation. In support of this claim, Figure~\ref{fig:motivation} illustrates a t-SNE~\cite{TSNE} bi-dimensional visualization of dense representations for the queries and the relevant documents of five manually rewritten conversations from the TREC 2019 CAsT dataset~\cite{DaltonEtAl2020}.
As illustrated, there is a clear spatial clustering among queries in the same conversation, as well as a clear spatial clustering of relevant documents for these queries.

We exploit 
locality to improve efficiency in conversational systems by caching the query results on the client side. 
Rather than caching pages of results answering queries likely to be resubmitted, we cache documents about a topic, believing that their content will be likewise relevant to successive queries issued by the user involved in the conversation.
Topic caching is effective in Web search~\cite{ipm2020-MTFP} 
but, as yet, was never explored in 
conversational search.


\ophir{Topic caching effectiveness rests on topical locality.  Specifically, if the variety of search domains is limited, the likelihood that past, and hence potentially cached, documents are relevant to successive searches is greater.  In the Web environment, search engines respond to a wide and diverse set of queries, and yet, topic caching is still effective \cite{ipm2020-MTFP}; thus, in the conversational search domain where a sequence of searches often focuses on a related if not on the same specific topic, topical caching, intuitively, should have even greater appeal than in the Web environment, motivating our exploration.}

To capitalize on the deep semantic relationship between conversation queries and documents, we leverage recent advances in Dense Retrieval (DR) models~\cite{ance,star,dpr,10.1145/3394486.3403305,CastDR}. In our DR setting,  documents are represented by low-dimension learned embeddings  stored for efficient access in a specialised metric index, such as that provided by the FAISS toolkit~\cite{JDH17}. Given a query embedded in the same multi-dimensional space, online ranking is performed by means of a top-$k$ nearest neighbor similarity search based on a metric distance. In the worst-case scenario, the computational cost of the nearest neighbor search is directly proportional to the number of documents stored in the metric index. 
To 
improve end-to-end responsiveness of the system, we insert a \emph{client-side} metric cache~\cite{Lucchese08, Lucchese12} in front of the DR system aimed at reusing documents retrieved for  previous queries in the same conversation. We investigate different strategies for  populating the cache at cold start and updating its content as the conversation topic evolves. 

Our metric cache returns an approximate result set for the current query.
Using reproducible experiments based on TREC \cast datasets, we demonstrate that our cache significantly reduces end-to-end conversational system processing times 
without answer quality degradation.
Typically, we answer a query without accessing the document 
index since the cache already stores the most similar documents.  More importantly, we can 
estimate the quality of the documents present in the cache for the current query, and based on this 
estimate, decide if querying the document index is potentially beneficial. 
Depending on the size of the cache, the hit rate measured on the \cast conversations varies between 65\% and 75\%, 
illustrating that caching significantly expedites conversational search by drastically reducing the number of queries submitted to the document index on the back-end.


Our contributions are as follows: 
\begin{itemize}
    \item Capitalizing on temporal locality, we propose a client-side document embedding cache \cache for expediting conversational search systems;
    \item We innovate means that assess current cache  content quality necessitating document index access only needed to improve response quality;
    \item Using the  TREC \cast datasets, we demonstrate responsiveness improvement without accuracy degradation.
\end{itemize}

The remainder of the paper is structured as follows:  Section~\ref{sec:architecture} introduces 
our conversational search system architecture and discusses   the proposed document embedding cache and the associated update strategies. Section~\ref{sec:expsetup} details our research questions, introducing the experimental settings and the experimental methodology.  Results  of our comprehensive evaluation conducted to answer the research questions are discussed in Section~\ref{sec:results}. Section~\ref{sec:related} contextualizes our contribution in the related work. Finally, we conclude our investigation in Section~\ref{sec:conclusions}.

\begin{figure}[htb]
\centering
\includegraphics[width=0.6\linewidth]{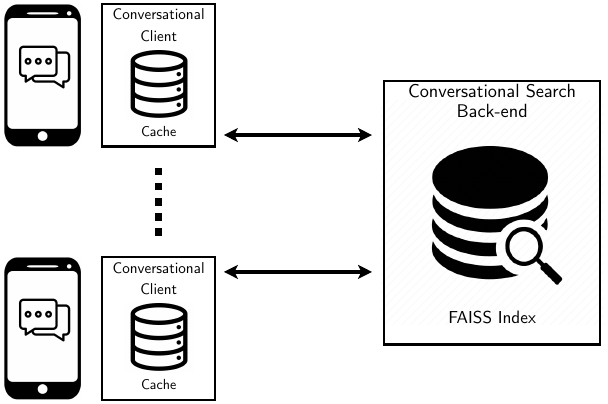}
\caption{Architecture of  a conversational search  system with client-side caching.}\label{fig:architecture}
\end{figure}

\section{A Conversational system with client-side caching} 
\label{sec:caching}
\label{sec:architecture}

A conversational search system enriched with our client-side caching is depicted in Figure \ref{fig:architecture}. We adopt a typical client-server architecture where a client supervises the conversational dialogue between a user and a search back-end running on a remote server.

We assume that the conversational back-end uses a dense retrieval model where documents and queries are both encoded with vector representations, also known as embeddings, in the same multi-dimensional latent space; the collection of document embeddings is stored, for efficient access, in a search system supporting nearest neighbor search, such as a FAISS index~\cite{JDH17}. 
Each conversational client, possibly running on a mobile device, deals with a single user conversation at a time, and hosts a local cache aimed at reusing, for efficiency reasons, the documents previously retrieved from the back-end as a result of the previous utterances of the ongoing conversation. Reusing previously retrieved, namely cached, results eliminates the additional index access, reducing latency and resource load. Specifically, the twofold goal of the cache is: 1) to improve  user-perceived responsiveness of the system by promptly answering user utterances with locally cached content; 2) to reduce the computational load on the back-end server by lowering the number of server requests as compared to an analogous solution not adopting client-side caching.

In detail, 
the client handles the  user conversation by  semantically enriching those utterances that lack context \cite{Mele2021AdaptiveUR} and encoding the rewritten utterance in the embedding space. Online conversational search is performed in the above settings by means of top $k$ nearest neighbor queries  based  on a metric distance between the embedding of the utterance and those of the indexed documents. The conversational client 
likewise queries the local cache or the back-end for the most relevant results answering the current  utterance and presents them to the requesting user.
\raf{The first query of a conversation is always answered by querying the back-end index, and the results retrieved are used to populate the initially empty cache. For successive utterances of the same conversation, the decision of whether to answer by leveraging the content of the cache or querying the remote index is taken locally as explained later.}
We begin by introducing the notation used, continuing with a mathematical background on the metric properties of queries and documents and with a detailed specification of our client-side cache together with an update policy based on the metric properties of query and document embeddings.
 
\subsection{Preliminaries}
Each query or document is represented by a vector in $\mathbb R^l$, hereinafter called an \textit{embedding}.
Let $\db = \{d_1,d_2,\ldots,d_n\}$  be a collection of $n$ documents represented by the embeddings $\Phi = \{\phi_1, \phi_2, \ldots, \phi_n\}$, where $\phi_i = \mathcal{L}(d_i)$ and $\mathcal{L}: \mathcal{D} \to \mathbb{R}^l$ is a learned representation function. Similarly, let  $q_a$ be a query represented by the embedding $\psi_a = \mathcal{L}(q_a)$ in the same multi-dimensional space $\mathbb{R}^l$. 

Similarity functions 
to compare embeddings exist including inner product~\cite{ance,dpr,star, sbert} and euclidean norm~\cite{colbert}.
We use STAR \cite{star} to encode queries and documents. Since STAR embeddings are fine-tuned for  maximal inner-product search, they cannot natively exploit  the plethora of efficient algorithms developed for searching in Euclidean metric spaces.

To leverage nearest neighbor search and all the efficient tools devised for it, maximum inner product similarity search between embeddings can be adapted to use the Euclidean distance. Given a query embedding $\psi_a \in \mathbb{R}^l$ and a set of document embeddings $\Phi = \{\phi_i\}$ with $\phi_i \in \mathbb{R}^l$, we apply the following transformation from  $\mathbb{R}^l$ to $\mathbb{R}^{l+1}$~\cite{10.5555/3045118.3045323,xbox}:
\begin{equation}
\label{eq:transformation}
    \bar{\psi}_a = \begin{bmatrix}\psi_a^T/\|\psi_a\| & 0\end{bmatrix}^T,\quad
    \bar{\phi}_i = \begin{bmatrix}\phi_i^T/M & \sqrt{1 - \|\phi_i\|^2/M^2}\end{bmatrix}^T,
\end{equation}
where $M = \max_i \|\phi_i\|$. In doing so, the maximization problem of the inner product $\langle\psi_a,\phi_i\rangle$ becomes exactly  equivalent to the minimization problem of the Euclidean distance $\|\bar{\psi}_a - \bar{\phi}_i\|$. In fact,
we have:
\begin{equation*}
     \min \|\bar{\psi}_a - \bar{\phi}_i\|^2 =
     \min \big( \|\bar{\psi}_a\|^2 + \|\bar{\phi}_i\|^2 - 2 \langle \bar{\psi}_a, \bar{\phi}_i \rangle \big) =
    \min \big(2 - 2\langle {\psi}_a, {\phi}_i/M \rangle \big) =
    \max \langle {\psi}_a, {\phi}_i\rangle.
\end{equation*}
Hence, hereinafter we consider the task of online ranking with a dense retriever  as a nearest neighbor search task based on the Euclidean distance among the transformed embeddings $\bar{\psi}$ and $\bar{\phi}$ in $\mathbb{R}^{l+1}$. 
Intuitively, assuming $l = 2$, the transformation~\eqref{eq:transformation} maps arbitrary \textit{query and document} vectors in $\mathbb{R}^2$ into unit-norm \textit{query and document} vectors in $\mathbb{R}^3$, i.e., the transformed vectors are mapped on the surface of the unit sphere in $\mathbb{R}^3$.

To simplify the notation we drop the bar symbol from the embeddings $\bar{\psi} \to \psi$ and $\bar{\phi} \to \phi$, and assume that the learned function $\mathcal{L}$ encodes queries and documents directly in $\mathbb{R}^{l+1}$ by also applying the above transformation.

\subsection{Nearest neighbor queries and metric distances}
Let $\delta$ be a {\em metric distance function}, $\delta: \mathbb{R}^{l+1} \times \mathbb{R}^{l+1} \to \mathbb{R}$,  measuring the Euclidean distance between two embeddings in $\mathbb{R}^{l+1}$ of valid documents and queries; 
the smaller the distance between the embeddings, the more similar the corresponding documents or queries are. 

Given a query $q_a$, we are interested in retrieving $\text{NN}(q_a, k)$, i.e., the $k$ Nearest Neighbor documents to $q_a$ query according to the distance function $\delta(\cdot, \cdot)$. 
In the metric space $\mathbb{R}^{l+1}$, $\text{NN}(q_a, k)$ identifies an hyperball $\mathcal{B}_a$ centered on $\psi_a = \mathcal{L}(q_a)$ and with radius $r_a$, computed as:
\begin{equation}\label{eq:radiusa}
r_a = \max_{d_i \in \text{NN}(q_a, k)} \delta(\psi_a, \mathcal{L}(d_i)).
\end{equation}
The radius $r_a$ is thus the distance from $q_a$ of the least similar document  among the ones in $\text{NN}(q_a, k)$\footnote{Without loss of generality, we assume that the least similar document is unique, and we do not have two or more documents at distance $r_a$ from $q_a$.}. 

\begin{figure}[htb!]
\centering
\includegraphics[width=0.4\linewidth]{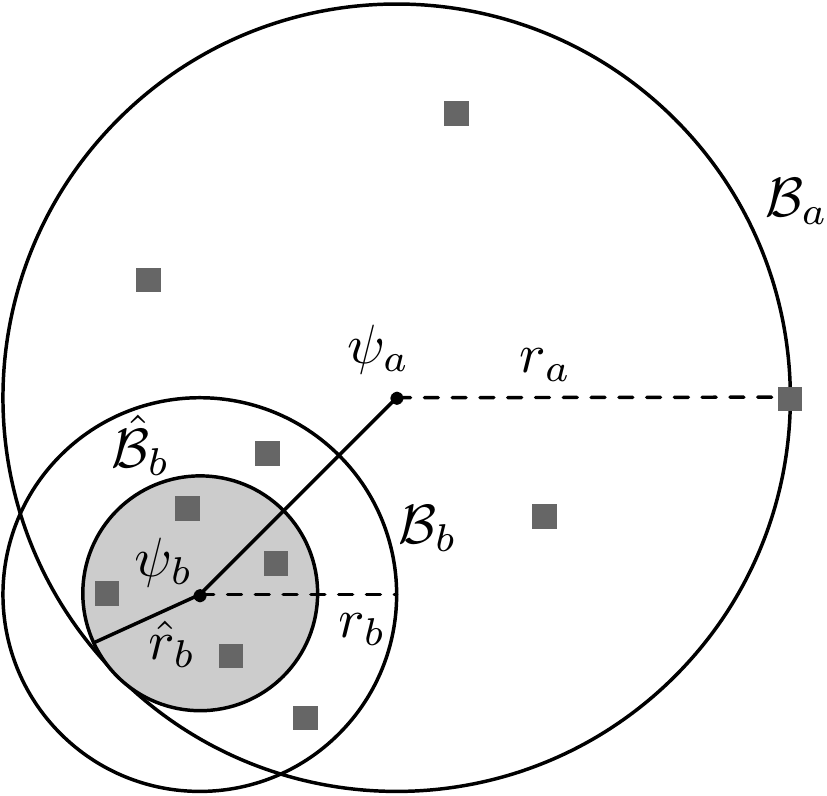}
\caption{Overlapping hyperballs for  $\text{NN}(q_a, 10)$ and $\text{NN}(q_b, 6)$  with embeddings in $\mathbb{R}^2$. Grey squares represent the embeddings of the 10 nearest neighbor documents to $q_a$.}\label{fig:overlap}
\end{figure}

We now introduce a new query $q_b$. Analogously, the set $\text{NN}(q_b, k)$  identifies  the  hyperball $\mathcal{B}_b$  with radius $r_b$ centered in $\psi_b$ and including the $k$ embeddings closest to $\psi_b$. If $\psi_a \neq \psi_b$, the two hyperballs can be completely disjoint, or may partially overlap. We introduce the quantity:
\begin{equation}\label{eq:radiusb}
\hat{r}_b = r_a -  \delta(\psi_a, \psi_b),
\end{equation}
to detect the case of a partial overlap in which the query embedding $\psi_b$ falls within the hyperball 
$\mathcal{B}_a$, i.e., $\delta(\psi_a, \psi_b) < r_a$, or, equivalently, $\hat{r}_b > 0$, as illustrated\footnote{The figure approximates the metric properties in a local neighborhood of $\psi_a$ on the $(l+1)$-dimensional unit sphere, i.e., in its locally-Euclidean $l$-dimensional tangent plane.} in Figure~\ref{fig:overlap}.

In this case, there always exists a hyperball $\hat{\mathcal{B}}_b$, centered on $\psi_b$ with radius $\hat{r}_b$ such that $\hat{\mathcal{B}}_b \subset \mathcal{B}_a$. As shown in the figure, some of the documents in $\text{NN}(q_a, k)$, retrieved for query $q_a$, may belong also to $\text{NN}(q_b, k)$. Specifically, these documents are all those within the hyperball $\hat{\mathcal{B}}_b$. Note that there can be other documents in $\mathcal{B}_a$ whose embeddings are contained in $\mathcal{B}_b$, but if such embeddings are in $\hat{\mathcal{B}}_b$, we have the \textit{guarantee} that the corresponding documents are the most similar to $q_b$ among \textcolor{black}{all the documents in \db~\cite{Lucchese08}.
Our experiments will show that the documents relevant for successive queries in a conversation overlap significantly.
To take advantage of such overlap,} we now introduce a cache for storing  historical embeddings that exploits the  above metric properties of dense representations of queries and documents. \raf{Given the representation on the current utterance, the proposed cache aims at reusing the embeddings already retrieved for previous  utterances of the same conversation  for improving the responsiveness of the system. In the simplistic example depicted in Figure \ref{fig:rb_hat}, our cache would answer query $q_b$ by reusing the embeddings in $\mathcal{B}_b$ already retrieved for $q_a$.}

\subsection{A metric cache for conversational search}

Since several queries in a multi-turn conversation may deal with the same broad topic, documents retrieved for the starting topic of a conversation might become useful also for answering subsequent queries within the same conversation. 
The  properties of nearest neighbor queries in metric spaces discussed in the previous subsection suggest a simple, but effective way to exploit temporal locality by means of a metric cache \cache deployed on the client-side of a conversational DR system. 

\begin{algorithm2e}[htb!]
	\DontPrintSemicolon
	\SetKwInOut{Input}{Input}\SetKwInOut{Output}{Output}

	\SetKwProg{myalg}{}{:}{end}
	\SetKwFunction{name}{{\sc }}

	\Input{a metric index \mindex, a metric cache \cache, a \raf{query}  cutoff $k$, a~cache~cutoff~$k_c$, a query embedding $\psi$}
	\Output{a results set \results}
		{
	 
    	\nl \If{{\sf Empty}$(\cache)$ {\bf or} {\sf LowQuality($\psi$, \cache)}}{
            \nl \results $\leftarrow {\sf NN}(\mindex, \psi, k_c)$\;
            \nl {\sf Insert}$(\cache, \results)$\;
            }
	    \nl \results $\leftarrow$ {\sf NN}$(\cache,\psi,k)$\;
	    \nl \Return \results\;
	    }
	\caption{The \cachealg\ pseudo-code}
	\label{algo:cache}
\end{algorithm2e}

Our system for CAChing Historical Embeddings (\cachealg) is specified in Algorithm \ref{algo:cache}. The system receives a sequence of queries belonging to a user conversation and answers them returning $k$ documents retrieved from the metric cache \cache 
or the metric index \mindex containing the document embeddings of the whole collection.  

When the conversation is initiated with a query $q$, whose embedding is $\psi$, the cache is empty (line 1). The main index \mindex, \raf{possibly stored on a remote back-end server,} is thus queried for top $\textsf{NN}(\mindex, \psi, k_c)$ documents, with  cache cutoff $k_c \gg k$ (line 2). Those $k_c$ documents are then stored in the cache (line 3).
The rationale of using a cache cutoff $k_c$ much larger than the \raf{query cutoff} $k$ is that of filling the cache with documents that are likely to be relevant also for the successive queries of the conversation, i.e., possibly all the documents in the conversation clusters depicted in Figure~\ref{fig:motivation}. The cache cutoff $k_c$ relates in fact with the radius $r_a$ of the hyperball $\mathcal{B}_a$ illustrated in Figure~\ref{fig:overlap}: the larger $k_c$  the larger $r_a$ and the possibility of having documents relevant to the successive queries of the conversation in the  hyperball $\mathcal{B}_a$. 
When a new query of the same conversation arrives, we estimate the quality of the historical embeddings stored in the cache for answering it. This is accomplished by the function {\sf LowQuality}($\psi,\cache)$  (line 1). If the results available in the cache \cache are likely to be of low quality, we issue the query to the main index \mindex with cache cutoff $k_c$ and add the top $k_c$ results to \cache (line 2-3).
Eventually, we query the cache for the $k$ nearest neighbor documents (line 4), and return them (line 5).

\subsubsection*{Cache quality estimation}
The  quality of the  historical embeddings  stored in \cache for answering a new query is estimated  heuristically within the function {\sf LowQuality}($\psi,\cache)$ called in line 1 of Algorithm \ref{algo:cache}.
Given the embedding $\psi$ of the new query, we first identify the query embedding $\psi_a$ closest to $\psi$ among the ones present in \cache, i.e., 
\begin{equation}\label{eq:minq}
\psi_a = \argmin_{\psi_i \in \cache} \delta(\psi_i, \psi)
\end{equation}
Once $\psi_a$ is identified, we consider the radius  $r_a$ of the hyperball $\mathcal{B}_a$, depicted in Figure \ref{fig:motivation}, and use Eq. \ref{eq:radiusb} to check if $\psi$ falls within $\mathcal{B}_a$. If this happen, it is likely that some of the documents  previously retrieved  for $\psi_a$ and stored in \cache are relevant even for $\psi$. 
Specifically, our quality estimation heuristics considers the value  $\hat{r} = r_a - \delta(\psi_a, \psi)$ introduced in Eq. \ref{eq:radiusb}. If $ \hat{r} > \epsilon$, with $\epsilon \geq 0$ being a hyperparameter of the cache, we answer $\psi$ with the $k$ nearest neighbor documents stored in the cache, i.e., the {\sf NN}$(\cache,\psi, k)$ documents; otherwise, we query the main embedding index in the conversational search back-end and update the cache accordingly.
This quality test has the advantage of efficiency; it simply requires computing
the distances between  $\psi$ and the embeddings of the few queries previously used to populate the cache for the current conversation, i.e., the ones that caused a cache miss and were answered by retrieving the embeddings from the back-end (lines 2 and 3 of Algorithm \ref{algo:cache}).

In addition, by changing the single hyperparameter $\epsilon$ that measures the distance of a query from the internal border of the hyperball containing the closest cached query, we can easily tune the quality-assessment heuristic for the specific needs.
In the experimental section, we propose and discuss a simple but effective technique for tuning $\epsilon$ to 
balance the effectiveness of the results returned and the efficiency improvement introduced with caching.


\section{Research questions and Experimental Settings}\label{sec:expsetup}

We now  present the research questions and the experimental setup aimed at evaluating the proposed \cachealg\ system in operational scenarios. 
That is, we experimentally assess both the accuracy, namely not hindering response quality, and efficiency, namely a reduction of index request time, of a conversational search system that includes \cachealg.  Our reference baseline is exactly the same conversational search system illustrated in Figure \ref{fig:architecture} where conversational clients always forward the queries to the back-end server managing the document embedding index.

\subsection{Research Questions}
Specifically, in the following we address the following research questions:
\begin{itemize}[align=left,leftmargin=*]
    \item \textbf{RQ1}: Does \cachealg\ provide effective answers to conversational utterances by reusing the embeddings retrieved for previous utterance of the same conversation? 
    \begin{itemize}
        \item[A.] How effective is the quality assessment heuristic used to decide cache updates? 
        \item[B.] To what extent does \cachealg\ impact on client-server interactions?  
        \item[C.] How much memory \cachealg\ requires in the worst case?
    \end{itemize}
    \item \textbf{RQ2}: How much does \cachealg\ expedite the conversational search process?   
    \begin{itemize}
        \item[A.] What is the impact of the cache cutoff $k_c$ on the efficiency of the system in case of cache misses?
        \item[B.] How much faster is answering a query from the cache rather than from the remote index?
    \end{itemize}
\end{itemize}




\subsection{Experimental settings}\label{ssec:expsetup}
Our conversational search system uses STAR~\cite{star} to encode \cast queries and documents as embeddings with 769 dimensions\footnote{STAR encoding uses 768 values but we added one dimension to each embedding  by applying the transformation in Eq. \ref{eq:transformation}.}. The document embeddings are stored in a dense retrieval system leveraging the FAISS library~\cite{JDH17} to efficiently perform similarity searches between queries and documents. The nearest neighbor search is exact, and no approximation/quantization mechanisms are deployed.

\paragraph{Datasets and dense representation}

Our experiments are based on the resources provided by the 2019, 2020, and 2021 editions of the TREC  Conversational Assistant Track (\cast). 
The \cast 2019 dataset consists of 50 human-assessed conversations, while the other two datasets include 25 conversations each, with an average of 10 turns per conversation. The \cast 2019 and 2020 include relevance judgements at passage level, whereas for \cast 2021 the relevance judgments are provided at the document level. 
The judgments, graded on a three-point scale,  refer to passages  of the TREC CAR (Complex Answer Retrieval),
and \msmarco (MAchine Reading COmprehension) collections for \cast 2019 and 2020, and to documents of \msmarco, KILT, Wikipedia, and Washington Post 2020 for \cast 2021\footnote{\url{https://www.treccast.ai/}}.

\raf{Regarding the dense representation of queries and passages/documents, our caching strategy is orthogonal w.r.t. the choice of the embedding. The state-of-the-art  single-representation models proposed in the literature are: DPR~\cite{dpr}, ANCE~\cite{ance}, and STAR~\cite{star}. The main difference among these models is how the fine-tuning of the underlying pre-trained language model, i.e., BERT, is carried out. We selected for our experiments the embeddings computed by the STAR model since it employs hard negative sampling during fine-tuning, obtaining better representations in terms of effectiveness w.r.t. ANCE and DPR.}
For  \cast 2019 and 2020, we generated a STAR embedding for each passage in the collections, while for \cast 2021, we encoded each document, up to the maximum input length of 512 tokens, in a single STAR embedding. 

Given  our  focus on the efficiency of conversational search, we strictly use manually rewritten queries, where missing keywords or mentions to previous subjects, e.g., pronouns, are resolved by human assessors.

\paragraph{\cachealg\ Configurations}
To answer our research questions, we measure 
 the end-to-end performance of the proposed \cachealg\ system on the three \cast datasets.
We compare \cachealg\ against the efficiency and effectiveness of a \textit{baseline} conversational search system with no caching, always answering  the conversational queries by using the  FAISS index hosted by the back-end (hereinafter indicated as \textit{no-caching}). The effectiveness of no-caching on the assessed conversations of the three \cast datasets represents an upper bound for the effectiveness of our \cachealg system. Analogously, we consider the  no-caching baseline always retrieving documents via the back-end as a lower bound for the responsiveness of the  conversational search task addressed.

We experiment with two different versions of our \cachealg system:
\begin{itemize}
    \item a \textit{static}-\cachealg: a metric cache populated with the $k_c$ nearest documents returned by the index for the first query of each conversation and never updated for the remaining queries of the conversations;
    \item a \textit{dynamic}-\cachealg: a metric cache updated at query processing time according to Alg.~\ref{algo:cache}, where  {\sf LowQuality}($\psi_b,\cache$) returns false if $\hat{r}_b \geq \epsilon$ (see Eq. \ref{eq:radiusb}) for at least one of the previously cached queries, and true otherwise.
\end{itemize}
We vary the cache cutoff $k_c$ in $\{1K, 2K, 5K, 10K\}$ and assess its impact. Additionally,
since conversations are typically brief, e.g., from 6 to 13 queries for the three \cast datasets considered, for efficiency and simplicity of design, we forgo implementing any space-freeing, eviction policy should  
the client-side cache reach maximum capacity. 
We assess experimentally that, even without eviction, the amount of memory needed by our \textit{dynamic}-\cachealg  to store the embeddings of the documents retrieved from the FAISS index during a single conversation suffices and does not present an issue.
In addition to the document embeddings, we recall that to implement the {\sf LowQuality}$(\cdot,\cdot)$ test, our cache records also the embeddings $\psi_a$ and radius $r_a$ of all the previous  queries $q_a$ of the conversation answered on the back-end.

\paragraph{Effectiveness Evaluation}
\looseness -1 The effectiveness of the \textit{no-caching} system, the static-\cachealg, and the dynamic-\cachealg\ are assessed by using the official metrics used to evaluate \cast conversational search systems~\cite{DaltonEtAl2020}: mean average precision at \raf{query} cutoff 200 (MAP@200), mean reciprocal rank at \raf{query} cutoff 200 (MRR@200), normalized discounted cumulative gain at \raf{query} cutoff 3 (nDCG@3), and precision at \raf{query}  cutoffs 1 and 3 (P@1, P@3). 
Our experiments report the statistically significant differences w.r.t. the baseline system for $p<0.01$ according to the two-sample t-test.

In addition to these standard IR measures, we introduce a new metric to measure the quality of the approximate answers retrieved from the cache w.r.t. the correct results retrieved form the FAISS index. We define the \textit{coverage} of a query $q$ w.r.t. a cache \cache and a given \raf {query cutoff}  value $k$, as the intersection, in terms of nearest neighbor documents, between the top $k$ elements retrieved for the cache \cache and the exact top $k$ elements retrieved from the whole index $\mathcal{M}$, divided by $k$:
\begin{equation}\label{eq:acov}
    \acov_k(q) = \frac{|{\sf NN}(\cache, \psi, k) \cap {\sf NN}(\mindex, \psi, k)|}{k},
\end{equation}
where $\psi$ is the embedding of  query $q$. We report on the quality of the approximate answers retrieved from the cache by measuring the coverage $\acov_k$, averaged over the different queries. 
\raf{The higher $\acov_k$ at a given query cutoff $k$ is, greater is the quality of the approximate $k$ nearest neighbor  documents retrieved from the cache. Of course $\acov_k(q) = 1$ for a given cutoff $k$ and query $q$ means that  we retrieve from the cache or the main index exactly the same set of answers. Moreover,  these answers come out to be ranked in the same order by the  distance function adopted. Besides measuring the quality of the answers retrieved from the cache vs the main index, we use } 
the  metric $\acov_k$ also to tune the hyperparameter $\epsilon$. 

\begin{figure}[bt!]
\centering
\includegraphics[width=0.6\linewidth]{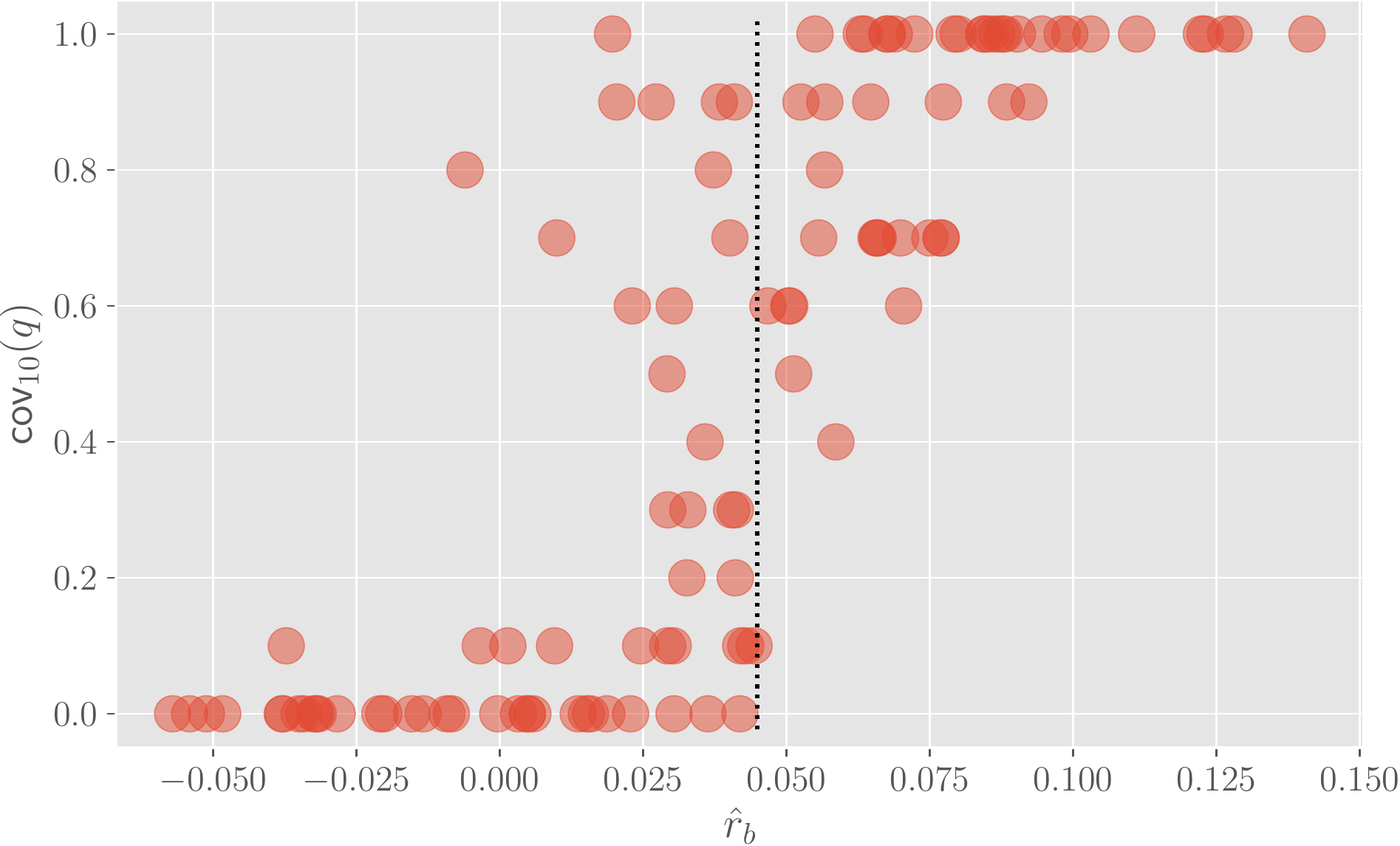}
\caption{Correlation  between $\hat{r}_b$ vs. \acov$_{10}(q)$ for the \cast 2019 train queries, using static-\cachealg, $k=10$ and $k_c=1K$. The vertical black dashed line corresponds to $\hat{r}_b = 0.04$, the tuned cache update threshold value $\epsilon$ used in our experiments.}\label{fig:rb_hat}
\end{figure}

To this end, Figure~\ref{fig:rb_hat} reports the correlation  between $\hat{r}_b$ vs. \acov$_{10}(q)$ for the \cast 2019 train queries, using static-\cachealg and $k_c=1K$. The queries with $\acov_{10} \leq 0.3$, i.e., those with no more than three documents in the intersection between the static-\cachealg contents and their actual top 10 documents, correspond to $\hat{r}_b \leq 0.04$. Hence, in our \raf{initial} experiments, we set the value of $\epsilon$  to $0.04$ \raf{ to obtain good coverage figures at small query cutoffs}. \raf{In answering RQ1.A we will also discuss  a different tuning of $\epsilon$  aimed at improving the effectiveness of dynamic-\cachealg at large query cutoffs. }

\paragraph{Efficiency Evaluation}
The efficiency of our \cachealg\ systems is measured in terms of: i) \textit{hit rate}, i.e., the percentage of queries, over the total number of queries, answered directly by the cache without querying the dense index; ii) \textit{average query response time} for our  \cachealg configurations and the \textit{no-caching} baseline.
The hit rate is measured by not considering the first query in each conversation since each conversation starts with an empty cache, and the first queries are thus compulsory cache misses, always answered by the index.
Finally, the query response time, namely latency, is measured as the amount of time from when a query is submitted to the system to the time it takes for the response to get back. To better understand the impact of caching, for \cachealg we measure separately the average response time for hits and misses.
The efficiency evaluation is conducted on a server equipped with an Intel Xeon E5-2630 v3 CPU clocked at 2.40GHz and 192 GiB of RAM. In our tests, we employ the FAISS\footnote{https://github.com/facebookresearch/faiss} Python API v1.6.4.  The experiments measuring query response time are conducted by using the low-level C++ exhaustive nearest-neighbor search FAISS APIs. 
We perform this choice to avoid possible overheads introduced by the Python interpreter that comes into play when using the standard FAISS high-level APIs. Moreover, as FAISS is a library designed and optimized for batch retrieval, our efficiency experiments are conducted by retrieving results for a batch of queries instead of a single one. The rationale of doing this relies in the fact that, on a back-end level, we can easily assume that queries coming from different clients can be batched together before being submitted to FAISS. The reported response times are obtained as an average of three different runs.

\paragraph{Available Software}
The source code used in our experiments is made publicly available to allow the reproducibility of the results\footnote{\url{https://github.com/hpclab/caching-conversational-search}}.

\section{Experimental Results}
\label{sec:results}

We now discuss the results of the experiments conducted to answer the research questions posed in Section \ref{sec:expsetup}.

\begin{table*}[htb!]
\centering
\caption{Retrieval performance measured on \cast datasets with or without document embedding caching. We highlight with symbol $\blacktriangledown$ statistical significant differences w.r.t.  \textit{no-caching}   for $p<0.01$ according to the two-sample t-test. Best values for each dataset and metric are shown in bold.}\label{tb:effectiveness}
\begin{adjustbox}{max width=\textwidth}
\begin{tabular}{cccccccccc}
\toprule
&& $k_c$ & {MAP@200} & {MRR@200} & {nDCG@3} & {P@1} & {P@3} & $cov_{10}$ & {Hit Rate} \\

\midrule
\multirow{9}{*}{\bf \cast 2019}&no-caching & -- &  \textbf{0.194} &	0.647 &	0.376 & 0.497 & 0.495 & -- &--\\
\cmidrule{3-10}
&\multirow{4}{*}{static-\cachealg}
 &  1K & 0.101$\blacktriangledown$ & 0.507$\blacktriangledown$ &  0.269$\blacktriangledown$ & 0.387$\blacktriangledown$ &  0.364$\blacktriangledown$ & 0.40 & 100\%  \\
 &&  2K & 0.112$\blacktriangledown$ & 0.567$\blacktriangledown$ & 0.304$\blacktriangledown$  &0.428  &0.414$\blacktriangledown$ & 0.47 &100\%  \\
 &&  5K & 0.129$\blacktriangledown$  & 0.588 & 0.316$\blacktriangledown$  & 0.451  & 0.426$\blacktriangledown$ & 0.56 & 100\% \\
 && 10K & 0.140$\blacktriangledown$  &0.611 & 0.338  & 0.486  & 0.459 & 0.62 & 100\% \\
\cmidrule{3-10}
&\multirow{4}{*}{dynamic-\cachealg}
 &  1K & 0.180$\blacktriangledown$ & 0.634 & 0.365  &  0.474 &  0.482 & 0.91 & 67.82\% \\
 &&  2K & 0.183$\blacktriangledown$	 & 0.631 &  0.366 & 0.480 & 0.487 & 0.93 & 70.69\% \\
 &&  5K & 0.186$\blacktriangledown$	 & 0.652 & 0.375  & 0.503  & 0.499 & 0.94 & 74.14\% \\
 && 10K & 0.190 & \textbf{0.655} & \textbf{0.380}  & \textbf{0.509}  &  \textbf{0.505} & 0.96 & 75.29\% \\
 \midrule
\multirow{9}{*}{\bf \cast 2020}&
no-caching & -- &  \textbf{0.212} & 0.622	 &	0.338 & 0.471 & 0.473 & -- & --\\	
\cmidrule{3-10}
&\multirow{4}{*}{static-\cachealg}
 &  1K &  0.112$\blacktriangledown$ & 0.421$\blacktriangledown$ &  0.215$\blacktriangledown$  & 0.312$\blacktriangledown$ &  0.306$\blacktriangledown$ & 0.35 & 100\%  \\
& &  2K &  0.120$\blacktriangledown$ & 0.454$\blacktriangledown$ &  0.236$\blacktriangledown$  & 0.351$\blacktriangledown$ &  0.324$\blacktriangledown$ & 0.41 & 100\%  \\
& &  5K &  0.139$\blacktriangledown$  & 0.509$\blacktriangledown$  &  0.267$\blacktriangledown$  & 0.394  &  0.370$\blacktriangledown$ & 0.48 & 100\% \\
& & 10K &  0.146$\blacktriangledown$  & 0.518$\blacktriangledown$ &  0.270$\blacktriangledown$  & 0.394$\blacktriangledown$ & 0.380$\blacktriangledown$ &  0.52  & 100\% \\
\cmidrule{3-10}
&\multirow{4}{*}{dynamic-\cachealg}
&  1K & 0.204$\blacktriangledown$ & 0.624 & 0.339  & \textbf{0.481}  &  0.478 & 0.91 & 56.02\% \\
& &  2K & 0.203$\blacktriangledown$ & \textbf{0.625} &  0.336 &  \textbf{0.481} &  0.470 & 0.93 & 60.73\% \\
& &  5K & 0.208 & 0.622 &  \textbf{0.341} &  0.476 &  \textbf{0.479} & 0.94 & 62.83\% \\
& &  10K & 0.210 & \textbf{0.625} &  0.339  & 0.476  &  0.476 & 0.96 & 63.87\% \\
 \midrule
\multirow{9}{*}{\bf \cast 2021}&
no-caching & -- & \textbf{0.109}  &	0.584 &	\textbf{0.340} & 0.449 & \textbf{0.411} & -- & --\\	
\cmidrule{3-10}
&\multirow{4}{*}{static-\cachealg}
 &  1K &  0.068$\blacktriangledown$ & 0.430$\blacktriangledown$  &  0.226$\blacktriangledown$  & 0.323$\blacktriangledown$ &  0.283$\blacktriangledown$ & 0.38 & 100\%  \\
& &  2K & 0.072$\blacktriangledown$  & 0.461$\blacktriangledown$ &  0.240$\blacktriangledown$  & 0.348$\blacktriangledown$ &  0.300$\blacktriangledown$ & 0.42 & 100\%  \\
& &  5K &  0.079$\blacktriangledown$  & 0.508$\blacktriangledown$  &  0.270$\blacktriangledown$  &  0.386 &  0.338$\blacktriangledown$ & 0.51 & 100\% \\
& & 10K &  0.080$\blacktriangledown$  & 0.503$\blacktriangledown$ & 0.272$\blacktriangledown$   & 0.367$\blacktriangledown$ &  0.338$\blacktriangledown$ & 0.56 & 100\% \\
\cmidrule{3-10}
&\multirow{4}{*}{dynamic-\cachealg}
&  1K & 0.106 & 0.577 & 0.335  & 0.443  & 0.409 & 0.89 & 61.97\% \\
& &  2K & 0.107 & \textbf{0.585} & 0.338  & \textbf{0.456}  & \textbf{0.411}  & 0.91 & 63.38\% \\
& &  5K & 0.106 & 0.584 & 0.334  & 0.449  & 0.407 & 0.92 & 66.67\% \\
& &  10K & 0.107 & 0.584 &  0.336 & 0.449  & 0.409 & 0.94 & 67.61\% \\
  \bottomrule

 \end{tabular}
 \end{adjustbox}
 \end{table*}

\subsection{RQ1: Can we provide effective cached answers?}

The results of the experiments conducted on the three \cast datasets with the \textit{no-caching} baseline,  static-\cachealg, and  dynamic-\cachealg\ are reported in Table~\ref{tb:effectiveness}.
For each dataset, the static, and dynamic versions of \cachealg, we vary the value of the cache cutoff $k_c$ as discussed in Sec.~\ref{ssec:expsetup}, and highlight with symbol $\blacktriangledown$ the   statistical significant  differences (two-sample t-test with $p<0.01$) w.r.t. the \textit{no-caching} baseline. The best results for each dataset and effectiveness metric are shown in bold.

By looking at the figures in the table, we see that  static-\cachealg  returns  worse results than \textit{no-caching} for all the datasets, most of the metrics, and cache cutoffs $k_c$ considered. However, in a few cases, the differences are not statistically significant. For example, we observe that static-\cachealg on \cast 2019 with $k_c=10k$  does not statistically differ from  \textit{no-caching} for all metrics but MAP@200. The  reuse of the embeddings retrieved  for the first queries of \cast 2019 conversations  is thus so high that even the simple heuristic of statically caching the top $10k$ embeddings of the first query  allows to answer effectively the following queries without further interactions with the back-end.
As expected, we see that by increasing the number $k_c$ of statically cached embeddings from $1K$ to $10K$, we improve the quality for all datasets and metrics. Interestingly, we observe that static-\cachealg performs relatively better at small query cutoffs since in  column P@1 we have, for 5 times out of 12, results not statistically different from those of \textit{no-caching}.
We explain such behavior by observing again Figure \ref{fig:overlap}: when an incoming query $q_b$ is close to a previously cached one, i.e., $\hat{r}_b \geq 0$, it is likely that the relevant documents for $q_b$ present in the cache are those most similar to $q_b$ among all those in \db. The larger is query cutoff $k$, the lower is the probability of the least similar documents among the ones in {\sf NN}$(q_b, k)$ residing in the cache.

When considering  dynamic-\cachealg, based on the heuristic update policy discussed earlier, effectiveness improves remarkably. Independently of the dataset and the value of $k_c$, we achieve performance figures that are not statistically different from those measured with  \textit{no-caching} for all metrics but MAP@200. Indeed, the metrics measured at small query cutoffs result in some cases to be even slightly better than those of the baseline even if the improvements are not statistically significant: since the embeddings relevant for a conversation are tightly clustered,  retrieving them from the cache rather than from the whole index  in some case reduces  noise and provides higher accuracy.
MAP@200 is the only metrics for which some configurations of dynamic-\cachealg result to perform worse than \textit{no-caching}. This is motivated by the tuning of threshold $\epsilon$ performed by focusing on small query cutoffs, i.e., the ones commonly considered important for conversational search tasks \cite{DaltonEtAl2020}.

\paragraph{RQ1.A: Effectiveness of the quality assessment heuristic}
The performance exhibited by dynamic-\cachealg demonstrates that the quality assessment heuristic used to determine cache updates is highly effective. 
To further corroborate this claim, 
the $cov_{10}$ column of Table~\ref{tb:effectiveness} reports for static-\cachealg and dynamic-\cachealg the mean coverage  for $k=10$ measured by averaging Eq.~\eqref{eq:acov} over all the conversational queries in the datasets. We recall that this measure counts the cardinality of the intersection between the top $10$ elements retrieved from the cache and the exact top $10$ elements retrieved from the whole index, divided by $10$. While the $cov_{10}$ values  for static-\cachealg range between 0.35 to 0.62, justifying the quality degradation captured by the metrics reported in the table, with dynamic-\cachealg we measure values between 0.89 and 0.96,  showing that, consistently across different datasets and cache configurations, the update heuristics proposed successfully trigger  when the content of the cache needs  refreshing to answer a new topic introduced in the conversation.

\begin{figure}[htb]
\centering
\includegraphics[width=0.6\linewidth]{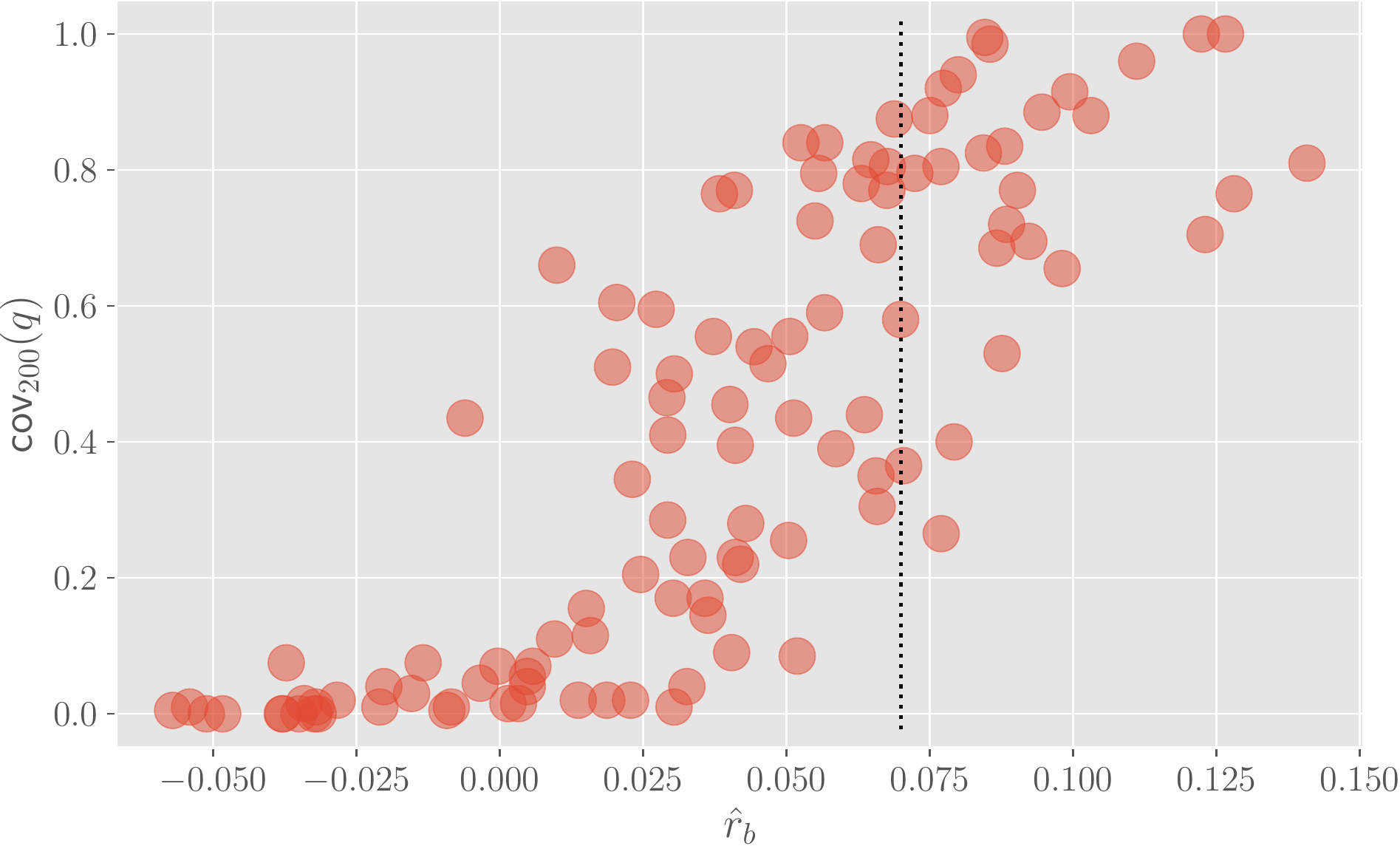}
\caption{Correlation  between $\hat{r}_b$ vs. \acov$_{200}(q)$ for the \cast 2019 train queries, using static-\cachealg and $k_c=1K$. The vertical black dashed line corresponds to $\hat{r}_b = 0.07$, the tuned cache update threshold value $\epsilon$ used in the experiments.
}\label{fig:rb_hat200}
\end{figure}

\raf{To gain further insights about RQ1.A, we conducted other experiments aimed at understanding if the  hyperparameter $\epsilon$ driving the dynamic-\cachealg updates can be fine-tuned for a specific query cutoff. Our investigation 
is motivated by
the MAP@200 results reported in Table \ref{tb:effectiveness} that are slightly lower than the baseline for $5$ out of $12$ dynamic-\cachealg configurations. We ask ourselves if it is possible to tune the value of $\epsilon$ to achieve MAP@200 results statistically equivalent to those of \textit{no-caching} without losing all the efficiency advantages of our client-side cache.}

\raf{Similarly to Figure \ref{fig:rb_hat},  the plot  in Figure \ref{fig:rb_hat200} 
shows the correlation between the value of $\hat{r}_b$ vs. \acov$_{200}(q)$ for the \cast 2019 train queries with static-\cachealg, $k=200$ and $k_c=1K$. Even at query cutoff $200$, we observe a strong correlation between $\hat{r}_b$ and the coverage metrics of Eq. \ref{eq:acov}: most of the train queries  with coverage $\acov_{200} \leq 0.3$ have a value of $\hat{r}_b$  smaller than $0.07$, with a single query for which this rule of thumb does not strictly hold. Hence, we set $\epsilon = 0.07$ and we run again our experiments with dynamic-\cachealg by varying the cache cutoff $k_c$ in $\{1k, 2k, 5k, 10k\}$.}
\raf{The results of these experiments, conducted with the \cast 2019 dataset, are reported in Table \ref{tb:effectiveness200}.
As we can see from the figures reported in the table,  increasing from $0.04$ to $0.07$ the value of $\epsilon$ improves the quality of the results returned by the cache at large cutoffs. Now dynamic-\cachealg returns results that are always, even for MAP@200, statistically equivalent to the ones retrieved from the whole index by the \textit{no-caching} baseline (according to a two-sample t-test for $p<0.01$). The improved quality at cutoff $200$ is of course paid with a decrease in efficiency. While for  $\epsilon = 0.04$ (see Table \ref{tb:effectiveness}) we measured on \cast 2019 hit rates ranging from $67.82$ to $75.29$, by setting $\epsilon = 0.07$ we strengthen the constraint on cache content quality and  correspondingly increase the number of cache updates performed. Consequently, the hit rate now ranges from $46.55$ to $58.05$, witnessing a likewise strong efficiency boost with respect to the \textit{no-caching} baseline. 
}

\begin{table*}[htb!]
\centering
\caption{Retrieval performance  on  \cast 2019  of the \textit{no-caching} baseline and  dynamic-\cachealg with $\epsilon = 0.07$. }\label{tb:effectiveness200}
\begin{adjustbox}{max width=\textwidth}
\begin{tabular}{ccccccccc}
\toprule
& $k_c$ & {MAP@200} & {MRR@200} & {nDCG@3} & {P@1} & {P@3} & $cov_{200}$ & {Hit Rate} \\
\midrule
no-caching & -- &  0.194 &	0.647 &	0.376 & 0.497 & 0.495 & -- &--\\
\cmidrule{2-9}
\multirow{4}{*}{dynamic-\cachealg}
 &  1K & 0.193& 0.645 &  0.374 & 0.497 &  0.491 & 0.83 & 46.55\%  \\
 &  2K & 0.193 & 0.644  & 0.375  & 0.497  &  0.493  & 0.91 &  51.15\%  \\
 &  5K & 0.194 &  0.645 &  0.375 &  0.497 &  0.493  & 0.93 &  54.02\%  \\
 &  10K & 0.194 & 0.648  &  0.375 & 0.497  &  0.493  & 0.94 &  58.05\%  \\
  \bottomrule
 \end{tabular}
 \end{adjustbox}
 \end{table*}

\paragraph{RQ1.B: Impact of \cachealg on client-server interactions} 

The last column of Table \ref{tb:effectiveness}  reports the cache hit rate, i.e., the percentage of conversational queries  over the total answered with the cached embeddings without interacting with the conversational search back-end. Of course,  static-\cachealg results in a trivial 100\% hit rate since all the queries in a conversation are answered with the embeddings initially retrieved for answering the first query. The lowest possible workload on the back-end is however paid with a significant performance drop with respect to the \textit{no-caching} baseline. 
With dynamic-\cachealg, instead, we achieve high hit rates with the optimal answer quality discussed earlier. As expected, the greater the value of $k_c$, the larger the number of cached embeddings and the higher the hit rate. With $k_c=1K$, hit rates range between $56.02\%$ to $67.82\%$, meaning that even with the lowest cache cutoff experimented more than half of the conversation queries in the 3 datasets are answered directly by the cache, without forwarding the query to the back-end. For $k_c=10K$, the hit rate value is in the interval $[63.87\%-75.29\%]$, with more than $3/4$ of the queries in the \cast 2019 dataset answered directly by the cache.
If we consider the hit rate as a measure correlated to the amount of temporal locality present in the \cast conversations, we highlight the highest locality present in  the 2019 dataset: on this dataset dynamic-\cachealg  with $k_c=1K$ achieves a hit rate higher that the ones measured for $k_c=10K$ configurations on \cast 2020 and 2021.

\paragraph{RQ1.C: Worst-case \cachealg memory requirements}    

\raf{The memory occupancy of static-\cachealg is limited, fixed and known in advance. The worst-case amount of memory required by dynamic-\cachealg depends instead on the value of $k_c$ and on the  number of cache updates performed during a conversation. The parameter $k_c$ establishes the number of embeddings added to the cache after every cache miss. Limiting the value of $k_c$ can be necessary to respect  memory constraints on the client hosting the cache. Anyway, the larger $k_c$ is, the greater the performance of dynamic-\cachealg thanks to the increased  likelihood that upcoming queries in the conversation will be answered directly, without querying the back-end index. In our experiments, we varied $k_c$ in $\{1K, 2K, 5K, 10K\}$ always obtaining optimal retrieval performances thanks to the effectiveness and robustness of the cache-update heuristic.}

Regarding the number of cache updates performed, we consider as exemplary cases the most difficult conversations for our caching strategy in the three \cast datasets, namely topic 77, topic 104, and topic 117 for \cast 2019, 2020, and 2021, respectively. These conversations require the highest number of cache updates: 6, 7, 6 for $k_c=1K$ and 5, 6, 5  for $k_c=10K$, respectively.
Consider topic 104 of \cast 2020, the toughest conversation for the memory requirements of  dynamic-\cachealg. At its maximum occupancy, after the last cache update,  dynamic-\cachealg system stores at most $8 \cdot 1K + 8 \approx 8K$ embeddings for $k_c=1K$ and $7 \cdot 1K + 7 \approx 70K$ embeddings for $k_c=10K$. In fact, at a given time, dynamic-\cachealg stores the $k_c$ embedding retrieved for the first query in the conversation plus  $k_c$ new embeddings for every cache update performed. Indeed, the total number is lower due to the presence of embeddings retrieved multiple times from the index on the back-end. The actual number of cache embeddings for the case considered is $7.5K$ and $64K$ for $k_c=1K$ and $k_c=10K$, respectively.
Since each embedding is represented with $769$ floating point values, the maximum memory occupation for our largest cache is  $64K \times 769 \times 4 $ bytes $\approx 188$ MB. Note that if we consider the case dynamic-\cachealg, $k_c=1K$, achieving  the same optimal performance of dynamic-\cachealg, $k_c=10K$ on \cast 2020 topic 104, the maximum occupancy of the cache decreases dramatically to about $28$ MB.

\subsection{RQ2: How much does \cachealg\ expedite the conversational search process?}
We now answer RQ2 by assessing the efficiency of the conversational search process in presence of cache misses (RQ2.A) or cache hits (RQ2.B).

\paragraph{RQ2.A: What is the impact of the cache cutoff $k_c$ on the efficiency of the system in case of cache misses?}
We first conduct experiments to understand the impact of  $k_c$ on the latency of nearest-neighbor queries performed on the remote back-end. To this end, we do not consider the costs of client-server communications, but only the retrieval time measured for answering a query on the remote index. Our aim is understanding if the value of $k_c$ impacts significantly or not the retrieval cost. In fact, when we answer the first query in the conversation or dynamic-\cachealg performs an update of the cache in case of a miss (lines 1-3 of Algorithm \ref{algo:cache}), we retrieve from the remote index a large set of $k_c$ embeddings to increase the likelihood of storing in the cache documents relevant for successive queries. However, the query cutoff $k$ commonly used for answering conversational queries is very small, e.g., $1, 3, 5$, and $k \ll k_c$. Our caching approach can improve efficiency only if the cost of retrieving from the remote index $k_c$ embeddings is comparable to that of retrieving a much smaller set of $k$ elements. Otherwise, even if we reduce remarkably the number of accesses to the back-end, every retrieval of a large number of results for filling or updating the cache would jeopardize its efficiency benefits.

We conduct the experiment on the \cast 2020 dataset by reporting the average latency (in msec.) of performing {\sf NN}$(q, k_c)$ queries on the remote index. Due to the peculiarities of the FAISS library implementation previously discussed, the response time is  measured by retrieving the top-$k_c$ results for a batch of $216$ queries, i.e., the \cast 2020 test utterances, and by averaging the total response time (Table \ref{tb:efficiency1}). Experimental results show that the  back-end query response time is approximately $1$ second and  is almost not  affected by the value of $k_c$. This is expected as exhaustive nearest-neighbor search requires the computation of the distances from the query of all indexed documents, plus the negligible cost of maintaining the top-$k_c$ closest documents in a min-heap. The result thus confirms that large $k_c$  values do not  jeopardize the efficiency of the whole system when cache misses occur.

\begin{table*}[h!]
\centering
\caption{Average  response time (msec.) for querying the FAISS back-end (no-caching) or the {static-\cachealg} and {dynamic-\cachealg} in case of cache hit.\label{tb:efficiency1}}
\begin{tabular}{ccccc}
\toprule
& \multicolumn{4}{c}{$k_c$} \\
\cmidrule{2-5}
& {1K} & {2K} & {5K} & {10K} \\
\midrule
no-caching  & 1,060 &1,058 & 1,061 & 1,073 \\
\midrule
{static-\cachealg}   & 0.14 & 0.30 & 0.78 & 1.59 \\
{dynamic-\cachealg}  & 0.36 & 0.70 & 1.73 & 3.48 \\
\bottomrule
\end{tabular}
\end{table*}

\paragraph{RQ2.B: How much faster is answering a query from the local cache rather than from the remote index?}
The second experiment conducted aims at measuring the average retrieval time for querying the client-side cache (line 4 of Algorithm \ref{algo:cache}) in case of  hit. We run the experiment for the two caches proposed, i.e., {static-\cachealg} and {dynamic-\cachealg}. While the first one stores a fixed number of documents, the latter employs cache updates that add document embeddings to the cache during the conversation. 
We report, in the last two rows of Table \ref{tb:efficiency1}, the average  response time of top-3 nearest-neighbor queries resulting in cache hits for different configurations of  {static-\cachealg} and {dynamic-\cachealg}. As before, latencies are measured on batches of $216$ queries, i.e., the \cast 2020 test utterances, by averaging the total response time. The results of the experiment show that, in case of a hit, querying the cache requires on average less than 4 milliseconds, more than 250 times less than querying the back-end. We observe that, as expected, hit time increases linearly with the size of the {static-\cachealg}. 
We also note that {dynamic-\cachealg} shows slightly higher latency than {static-\cachealg}. This is due to the updates of the cache performed during the conversation that add embeddings to the cache. 
This result shows that the use of a cache in conversational search allows to achieve a speedup of up to four orders of magnitude, i.e., from seconds to few tenths of milliseconds, between querying a remote index and a local cache.

We can now finally answer RQ2, how much does \cachealg\ expedite the conversational search process, by computing the average overall speedup achieved by our caching techniques on an entire conversation. Assuming that the average conversation is composed of $10$ utterances, the \textit{no-caching} baseline that always queries the back-end leads to a total response time of about $10 \times 1.06 = 10.6$ seconds. Instead, with {static-\cachealg} we perform only one retrieval from the remote index for the first utterance while the remaining queries are resolved by the cache. Assuming the use of {static-\cachealg} with 10K embeddings, i.e., the one with higher latency, the total response time for the whole  conversation is $1.06 + (9 \cdot 0.00159) = 1.074$ seconds, with an overall speedup of about $9.87\times$ over \textit{no-caching}. Finally, the use of {dynamic-\cachealg} implies possible cache updates that may increase the number of queries answered using the remote index. In detail, {dynamic-\cachealg} with 10K embeddings obtains a hit rate of about $64$\% on \cast 2020 (see Table \ref{tb:effectiveness}). This means that, on average, we forward $1 + (9 \cdot 0.36) = 4.24$ queries to the back-end that cost in total $4.24 \cdot 1.06 = 4.49$ seconds. The remaining cost comes from cache hits. Hits are on average $5.76$ and require $5.76 \cdot 0.00348 = 0.002$ seconds accounting for a total response time for the whole conversation of $4.242$ seconds. This leads to a  speedup of $2.5\times$ with respect to the \textit{no-caching} solution.

The above figures confirm the feasibility and the computational performance advantages of our client-server solution for caching historical embeddings for conversational search.

\section{Related Work}
\label{sec:related}

Our contribution relates to two main research areas. The first, attracting recently significant interest, is \textit{Conversational Search}.
Specifically, our work focuses on the ability of neural retrieval models to capture the semantic relationship between conversation utterances and documents, and, more centrally, with efficiency aspects of neural search. 
The second related area is \textit{Similarity Caching} that was initially investigated  in the field of content-based image retrieval  and  contextual advertisement.

\paragraph{Neural approaches for conversational search}
\fm{Conversational search focuses on retrieving relevant documents from a collection to fulfill user information needs expressed in a dialogue, i.e., sequences of natural-language utterances  expressed in oral or written form~\cite{https://doi.org/10.48550/arxiv.2201.05176,10.1145/3269206.3271776}.} Given the nature of speech, these queries often lack context and are grammatically poorly formed, complicating their processing. To address these issues, it is natural to exploit past queries and their system response, if available, in a given conversation to build up a \textit{context history}, and use this history to enrich the semantic contents of the current query. The context history is typically used to rewrite the query in a self-contained, decontextualized query, suitable for ad-hoc document retrieval~\cite{10.1145/3446426,Mele2021AdaptiveUR,Yang2019QueryAA,10.1145/3397271.3401130,10.1145/3498557}.
Lin \emph{et al.} propose two conversational query reformulation methods based on the combination of term importance estimation and neural query rewriting~\cite{10.1145/3446426}. For the latter, authors reformulate conversational queries into natural and human-understandable queries with a pretrained sequence-to-sequence model. They also use reciprocal rank fusion to combine the two approaches yielding state-of-the-art retrieval effectiveness in terms of NDCG@3 compared to the best submission of Text REtrieval Conference (TREC) Conversational Assistant Track (CAsT) 2019.
Similarly, Voskarides \emph{et al.} focus on multi-turn passage retrieval by proposing QuReTeC (Query Resolution by Term Classification), a neural query resolution model based on bidirectional transformers and a distant supervision method to automatically generate training data by using query-passage relevance labels~\cite{10.1145/3397271.3401130}. Authors incorporate QuReTeC in a multi-turn, multi-stage passage retrieval architecture and show its effectiveness on the TREC CAsT dataset.

Others approach the problem by leveraging pre-trained generative language model to directly generate the reformulated queries \cite{9413557,10.1145/3437963.3441748,10.1145/3397271.3401323}. Some other studies combine approaches based on term selection strategies and query generation methods \cite{kumar-callan-2020-making,10.1145/3446426}.
\fm{Xu \emph{et al.} propose to track the context history on a different level, i.e., by exploiting user-level historical conversations~\cite{xu-etal-2020-user}. They build a structured  per-user memory knowledge graph to represent users' past interactions and  manage current queries. The knowledge graph is dynamically updated and complemented with  a reasoning model that predicts optimal dialog policies to be used to build the personalized answers.}

Pre-trained language models, such as BERT~\cite{bert}, learn semantic representations called \emph{embeddings} from the contexts of words and, therefore, better capture the relevance of a document w.r.t.\ a query, with substantial improvements over the classical approach in the ranking and re-ranking of documents~\cite{lin2020pretrained}. 
Recently, several efforts exploited pre-trained language models to represent queries and documents in the same dense latent vector space, and then used inner product to compute the relevance score of a document w.r.t. a query.

In conversational search, the representation of a query can be computed in two different ways. In one case, a stand-alone contextual query understanding module reformulates the user query $q$ into a rewritten query $\hat{q}$, exploiting the context history $H$~\cite{https://doi.org/10.48550/arxiv.2201.05176}, and then a query embedding $\mathcal{L}(\hat{q})$ is computed. In the other case, the learned representation function is trained to receive as input the query $q$ together with its context history $H_q$, and to generate a query embedding $\mathcal{L}(q, H_q)$~\cite{convdr}.
In both cases, dense retrieval methods are used to compute the query-document similarity, by deploying efficient nearest neighbor techniques over specialised indexes, such as those provided by the FAISS toolkit~\cite{JDH17}.


\paragraph{Similarity caching}
Similarity caching is a variant of classical exact caching in which the cache can return items that are similar, but not necessarily identical, to those queried. Similarity caching  was first introduced by Falchi \emph{et al.}, where the authors proposed two caching algorithms possibly returning  approximate result sets for k-NN similarity queries \cite{Lucchese08}. The two caching algorithms differ in the strategies adopted for building the approximate result set and deciding its quality based on the properties of metric spaces discussed in Section \ref{sec:architecture}. The authors focused on large-scale content-based image retrieval and conducted tests on a collection of one million images observing a significant reduction in average response time. Specifically, with a cache storing at most 5\% of the total dataset, they achieved hit rates exceeding  20\%. In successive works, the same authors analyzed the impact of similarity caching on the retrieval from larger collections with real user queries \cite{LuccheseEDBT,Lucchese12}. 
Chierichetti \emph{et al.} propose a similar caching solution that is used to efficiently identify advertisement candidates on the basis of those retrieved for similar past queries \cite{Chierichetti09}. Finally, Neglia \emph{et al.} propose an interesting theoretical study of similarity caching in the offline, adversarial, and stochastic settings \cite{SimCache22}, aimed at understanding how to compute the expected cost of a given similarity caching policy.

\smallskip
We capitalize on these seminal works by exploiting the properties of similarity caching in metric spaces for a completely different scenario, i.e., dense retrievers for conversational search. Differently from image and advertisement retrieval, our use case is characterized by the similarity among successive queries in a conversation, enabling a novel solution based on integrating a small similarity cache in the conversational client. Our client-side similarity cache answers most of the queries in a conversation without querying the main index hosted remotely.
A similar work to our own is the one by Sermpezis \emph{et al.}, where authors propose a similarity-based system for recommending alternative cached content to a user when their exact request cannot be satisfied by the local cache \cite{SoftCache}. The contribution is related because it proposes a client-side cache where similar content is looked for, although their focus is on how statically fill the local caches on the basis of user profiles and content popularity.

\section{Conclusion}
\label{sec:conclusions}

We introduced a client-side, document-embedding cache for expediting conversational search systems. Although caching is extensively used in search, we take a closer look at how it can be effectively and efficiently exploited in a novel and challenging  setting: a client-server conversational architecture  exploiting state-of-the-art dense retrieval models and a novel metric cache hosted on the client-side. 

Given the high temporal locality of the embeddings retrieved for answering utterances in a conversation, a cache can provide a great advantage to expedite conversational systems. We initially prove that both queries and documents in a conversation lie close together in the embedding space and that given this specific interaction and query properties, we can exploit the metric properties of distance computations in a dense retrieval context.

We propose two types of caching and compare the results in terms of both effectiveness and efficiency with respect to a no-caching baseline using the same back-end search solution. The first is a static-\cachealg which populates the cache with documents retrieved based on the first query of a conversation only. The second, dynamic-\cachealg, proposes also an update mechanism that comes in place when we determine, via a precise and efficient heuristic strategy, that the current contents of the cache might not provide relevant results.

The results of extensive and reproducible experiments conducted on \cast datasets show that dynamic-\cachealg achieves hit rates up to 75\% with answers quality statistically equivalent  to that of the   \textit{no-caching} baseline. In terms of efficiency,  the response time varies with the size of the cache, nevertheless queries resulting in cache hit are three orders of magnitude faster  than  those processed on the back-end (accessed only for cache misses by dynamic-\cachealg and for all queries by the \textit{no-caching} baseline). 

We  conclude that our  \cachealg solution for conversational search is a viable and effective solution, also opening the door for significant further investigation. \fm{Its client-side organization permits, for example, to effectively integrate models of user-level contextual knowledge. Equally interesting is the investigation of user-level, personalized query rewriting strategies and neural representations.}

\begin{acks}
This work is supported by the European Union – Horizon 2020 Program under the scheme “INFRAIA-01-2018-2019 – Integrating Activities for Advanced Communities”, Grant Agreement n.871042, “SoBigData++: European Integrated Infrastructure for Social Mining and Big Data Analytics” (http://www.sobigdata.eu).
\end{acks}


\begin{thebibliography}{36}


\ifx \showCODEN    \undefined \def \showCODEN     #1{\unskip}     \fi
\ifx \showDOI      \undefined \def \showDOI       #1{#1}\fi
\ifx \showISBNx    \undefined \def \showISBNx     #1{\unskip}     \fi
\ifx \showISBNxiii \undefined \def \showISBNxiii  #1{\unskip}     \fi
\ifx \showISSN     \undefined \def \showISSN      #1{\unskip}     \fi
\ifx \showLCCN     \undefined \def \showLCCN      #1{\unskip}     \fi
\ifx \shownote     \undefined \def \shownote      #1{#1}          \fi
\ifx \showarticletitle \undefined \def \showarticletitle #1{#1}   \fi
\ifx \showURL      \undefined \def \showURL       {\relax}        \fi
\providecommand\bibfield[2]{#2}
\providecommand\bibinfo[2]{#2}
\providecommand\natexlab[1]{#1}
\providecommand\showeprint[2][]{arXiv:#2}

\bibitem[\protect\citeauthoryear{Anand, Cavedon, Joho, Sanderson, and
  Stein}{Anand et~al\mbox{.}}{2020}]%
        {anand2020conversational}
\bibfield{author}{\bibinfo{person}{Avishek Anand}, \bibinfo{person}{Lawrence
  Cavedon}, \bibinfo{person}{Hideo Joho}, \bibinfo{person}{Mark Sanderson},
  {and} \bibinfo{person}{Benno Stein}.} \bibinfo{year}{2020}\natexlab{}.
\newblock \showarticletitle{Conversational search}. In
  \bibinfo{booktitle}{\emph{Dagstuhl Reports}}, Vol.~\bibinfo{volume}{9}.
\newblock


\bibitem[\protect\citeauthoryear{Bachrach, Finkelstein, Gilad-Bachrach, Katzir,
  Koenigstein, Nice, and Paquet}{Bachrach et~al\mbox{.}}{2014}]%
        {xbox}
\bibfield{author}{\bibinfo{person}{Yoram Bachrach}, \bibinfo{person}{Yehuda
  Finkelstein}, \bibinfo{person}{Ran Gilad-Bachrach}, \bibinfo{person}{Liran
  Katzir}, \bibinfo{person}{Noam Koenigstein}, \bibinfo{person}{Nir Nice},
  {and} \bibinfo{person}{Ulrich Paquet}.} \bibinfo{year}{2014}\natexlab{}.
\newblock \showarticletitle{Speeding up the Xbox Recommender System Using a
  Euclidean Transformation for Inner-Product Spaces}. In
  \bibinfo{booktitle}{\emph{Proceedings of the 8th ACM Conference on
  Recommender Systems}} \emph{(\bibinfo{series}{RecSys '14})}.
  \bibinfo{publisher}{Association for Computing Machinery},
  \bibinfo{address}{New York, NY, USA}, \bibinfo{pages}{257–264}.
\newblock
\showISBNx{9781450326681}
\urldef\tempurl%
\url{https://doi.org/10.1145/2645710.2645741}
\showDOI{\tempurl}


\bibitem[\protect\citeauthoryear{Chierichetti, Kumar, and
  Vassilvitskii}{Chierichetti et~al\mbox{.}}{2009}]%
        {Chierichetti09}
\bibfield{author}{\bibinfo{person}{Flavio Chierichetti}, \bibinfo{person}{Ravi
  Kumar}, {and} \bibinfo{person}{Sergei Vassilvitskii}.}
  \bibinfo{year}{2009}\natexlab{}.
\newblock \showarticletitle{Similarity Caching}. In
  \bibinfo{booktitle}{\emph{Proceedings of the Twenty-Eighth ACM
  SIGMOD-SIGACT-SIGART Symposium on Principles of Database Systems}}
  \emph{(\bibinfo{series}{PODS '09})}. \bibinfo{publisher}{Association for
  Computing Machinery}, \bibinfo{address}{New York, NY, USA},
  \bibinfo{pages}{127–136}.
\newblock
\showISBNx{9781605585536}
\urldef\tempurl%
\url{https://doi.org/10.1145/1559795.1559815}
\showDOI{\tempurl}


\bibitem[\protect\citeauthoryear{Dalton, Xiong, Kumar, and Callan}{Dalton
  et~al\mbox{.}}{2020}]%
        {DaltonEtAl2020}
\bibfield{author}{\bibinfo{person}{Jeffrey Dalton}, \bibinfo{person}{Chenyan
  Xiong}, \bibinfo{person}{Vaibhav Kumar}, {and} \bibinfo{person}{Jamie
  Callan}.} \bibinfo{year}{2020}\natexlab{}.
\newblock \showarticletitle{CAsT-19: A Dataset for Conversational Information
  Seeking}. In \bibinfo{booktitle}{\emph{Proc. SIGIR}}.
  \bibinfo{pages}{1985–1988}.
\newblock


\bibitem[\protect\citeauthoryear{Devlin, Chang, Lee, and Toutanova}{Devlin
  et~al\mbox{.}}{2019}]%
        {bert}
\bibfield{author}{\bibinfo{person}{Jacob Devlin}, \bibinfo{person}{{Ming-Wei}
  Chang}, \bibinfo{person}{Kenton Lee}, {and} \bibinfo{person}{Kristina
  Toutanova}.} \bibinfo{year}{2019}\natexlab{}.
\newblock \showarticletitle{{BERT}: {P}re-training of {D}eep {B}idirectional
  {T}ransformers for {L}anguage {U}nderstanding}. In
  \bibinfo{booktitle}{\emph{Proc. {NAACL}}}.
\newblock


\bibitem[\protect\citeauthoryear{Falchi, Lucchese, Orlando, Perego, and
  Rabitti}{Falchi et~al\mbox{.}}{2008}]%
        {Lucchese08}
\bibfield{author}{\bibinfo{person}{Fabrizio Falchi}, \bibinfo{person}{Claudio
  Lucchese}, \bibinfo{person}{Salvatore Orlando}, \bibinfo{person}{Raffaele
  Perego}, {and} \bibinfo{person}{Fausto Rabitti}.}
  \bibinfo{year}{2008}\natexlab{}.
\newblock \showarticletitle{A Metric Cache for Similarity Search}. In
  \bibinfo{booktitle}{\emph{Proc. LSDS-IR}}. \bibinfo{pages}{43–50}.
\newblock


\bibitem[\protect\citeauthoryear{Falchi, Lucchese, Orlando, Perego, and
  Rabitti}{Falchi et~al\mbox{.}}{2009}]%
        {LuccheseEDBT}
\bibfield{author}{\bibinfo{person}{Fabrizio Falchi}, \bibinfo{person}{Claudio
  Lucchese}, \bibinfo{person}{Salvatore Orlando}, \bibinfo{person}{Raffaele
  Perego}, {and} \bibinfo{person}{Fausto Rabitti}.}
  \bibinfo{year}{2009}\natexlab{}.
\newblock \showarticletitle{Caching Content-Based Queries for Robust and
  Efficient Image Retrieval}. In \bibinfo{booktitle}{\emph{Proceedings of the
  12th International Conference on Extending Database Technology: Advances in
  Database Technology}} \emph{(\bibinfo{series}{EDBT '09})}.
  \bibinfo{publisher}{Association for Computing Machinery},
  \bibinfo{address}{New York, NY, USA}, \bibinfo{pages}{780–790}.
\newblock
\showISBNx{9781605584225}
\urldef\tempurl%
\url{https://doi.org/10.1145/1516360.1516450}
\showDOI{\tempurl}


\bibitem[\protect\citeauthoryear{Falchi, Lucchese, Orlando, Perego, and
  Rabitti}{Falchi et~al\mbox{.}}{2012}]%
        {Lucchese12}
\bibfield{author}{\bibinfo{person}{Fabrizio Falchi}, \bibinfo{person}{Claudio
  Lucchese}, \bibinfo{person}{Salvatore Orlando}, \bibinfo{person}{Raffaele
  Perego}, {and} \bibinfo{person}{Fausto Rabitti}.}
  \bibinfo{year}{2012}\natexlab{}.
\newblock \showarticletitle{Similarity Caching in Large-Scale Image Retrieval}.
\newblock \bibinfo{journal}{\emph{Inf. Process. Manage.}} \bibinfo{volume}{48},
  \bibinfo{number}{5} (\bibinfo{year}{2012}), \bibinfo{pages}{803–818}.
\newblock
\showISSN{0306-4573}


\bibitem[\protect\citeauthoryear{Gao, Xiong, Bennett, and Craswell}{Gao
  et~al\mbox{.}}{2022}]%
        {https://doi.org/10.48550/arxiv.2201.05176}
\bibfield{author}{\bibinfo{person}{Jianfeng Gao}, \bibinfo{person}{Chenyan
  Xiong}, \bibinfo{person}{Paul Bennett}, {and} \bibinfo{person}{Nick
  Craswell}.} \bibinfo{year}{2022}\natexlab{}.
\newblock \bibinfo{title}{Neural Approaches to Conversational Information
  Retrieval}.
\newblock
\newblock
\urldef\tempurl%
\url{https://doi.org/10.48550/ARXIV.2201.05176}
\showDOI{\tempurl}


\bibitem[\protect\citeauthoryear{Huang, Sharma, Sun, Xia, Zhang, Pronin,
  Padmanabhan, Ottaviano, and Yang}{Huang et~al\mbox{.}}{2020}]%
        {10.1145/3394486.3403305}
\bibfield{author}{\bibinfo{person}{Jui-Ting Huang}, \bibinfo{person}{Ashish
  Sharma}, \bibinfo{person}{Shuying Sun}, \bibinfo{person}{Li Xia},
  \bibinfo{person}{David Zhang}, \bibinfo{person}{Philip Pronin},
  \bibinfo{person}{Janani Padmanabhan}, \bibinfo{person}{Giuseppe Ottaviano},
  {and} \bibinfo{person}{Linjun Yang}.} \bibinfo{year}{2020}\natexlab{}.
\newblock \showarticletitle{Embedding-Based Retrieval in Facebook Search}. In
  \bibinfo{booktitle}{\emph{Proc. SIGKDD}}. \bibinfo{pages}{2553–2561}.
\newblock


\bibitem[\protect\citeauthoryear{Johnson, Douze, and Jegou}{Johnson
  et~al\mbox{.}}{2021}]%
        {JDH17}
\bibfield{author}{\bibinfo{person}{J. Johnson}, \bibinfo{person}{M. Douze},
  {and} \bibinfo{person}{H. Jegou}.} \bibinfo{year}{2021}\natexlab{}.
\newblock \showarticletitle{Billion-Scale Similarity Search with GPUs}.
\newblock \bibinfo{journal}{\emph{IEEE Trans. Big Data}} \bibinfo{volume}{7},
  \bibinfo{number}{03} (\bibinfo{year}{2021}), \bibinfo{pages}{535--547}.
\newblock


\bibitem[\protect\citeauthoryear{Karpukhin, Oguz, Min, Lewis, Wu, Edunov, Chen,
  and Yih}{Karpukhin et~al\mbox{.}}{2020}]%
        {dpr}
\bibfield{author}{\bibinfo{person}{Vladimir Karpukhin}, \bibinfo{person}{Barlas
  Oguz}, \bibinfo{person}{Sewon Min}, \bibinfo{person}{Patrick Lewis},
  \bibinfo{person}{Ledell Wu}, \bibinfo{person}{Sergey Edunov},
  \bibinfo{person}{Danqi Chen}, {and} \bibinfo{person}{Wen-tau Yih}.}
  \bibinfo{year}{2020}\natexlab{}.
\newblock \showarticletitle{Dense Passage Retrieval for Open-Domain Question
  Answering}. In \bibinfo{booktitle}{\emph{Proc. EMNLP}}.
  \bibinfo{pages}{6769--6781}.
\newblock


\bibitem[\protect\citeauthoryear{Khattab and Zaharia}{Khattab and
  Zaharia}{2020}]%
        {colbert}
\bibfield{author}{\bibinfo{person}{Omar Khattab} {and} \bibinfo{person}{Matei
  Zaharia}.} \bibinfo{year}{2020}\natexlab{}.
\newblock \showarticletitle{{ColBERT: Efficient and Effective Passage Search
  via Contextualized Late Interaction over BERT}}. In
  \bibinfo{booktitle}{\emph{Proc. SIGIR}}. \bibinfo{pages}{39–48}.
\newblock


\bibitem[\protect\citeauthoryear{Kumar and Callan}{Kumar and Callan}{2020}]%
        {kumar-callan-2020-making}
\bibfield{author}{\bibinfo{person}{Vaibhav Kumar} {and} \bibinfo{person}{Jamie
  Callan}.} \bibinfo{year}{2020}\natexlab{}.
\newblock \showarticletitle{Making Information Seeking Easier: An Improved
  Pipeline for Conversational Search}. In \bibinfo{booktitle}{\emph{Findings of
  the Association for Computational Linguistics: EMNLP 2020}}.
  \bibinfo{publisher}{Association for Computational Linguistics},
  \bibinfo{address}{Online}, \bibinfo{pages}{3971--3980}.
\newblock
\urldef\tempurl%
\url{https://doi.org/10.18653/v1/2020.findings-emnlp.354}
\showDOI{\tempurl}


\bibitem[\protect\citeauthoryear{Li, Li, and Nie}{Li et~al\mbox{.}}{2022}]%
        {10.1145/3498557}
\bibfield{author}{\bibinfo{person}{Yongqi Li}, \bibinfo{person}{Wenjie Li},
  {and} \bibinfo{person}{Liqiang Nie}.} \bibinfo{year}{2022}\natexlab{}.
\newblock \showarticletitle{Dynamic Graph Reasoning for Conversational
  Open-Domain Question Answering}.
\newblock \bibinfo{journal}{\emph{ACM Trans. Inf. Syst.}} \bibinfo{volume}{40},
  \bibinfo{number}{4}, Article \bibinfo{articleno}{82} (\bibinfo{date}{jan}
  \bibinfo{year}{2022}), \bibinfo{numpages}{24}~pages.
\newblock
\showISSN{1046-8188}
\urldef\tempurl%
\url{https://doi.org/10.1145/3498557}
\showDOI{\tempurl}


\bibitem[\protect\citeauthoryear{Lin, Nogueira, and Yates}{Lin
  et~al\mbox{.}}{2020}]%
        {lin2020pretrained}
\bibfield{author}{\bibinfo{person}{Jimmy Lin}, \bibinfo{person}{Rodrigo
  Nogueira}, {and} \bibinfo{person}{Andrew Yates}.}
  \bibinfo{year}{2020}\natexlab{}.
\newblock \bibinfo{title}{Pretrained Transformers for Text Ranking: {BERT} and
  Beyond}.
\newblock
\newblock
\showeprint[arxiv]{2010.06467}


\bibitem[\protect\citeauthoryear{Lin, Yang, Nogueira, Tsai, Wang, and Lin}{Lin
  et~al\mbox{.}}{2021}]%
        {10.1145/3446426}
\bibfield{author}{\bibinfo{person}{Sheng-Chieh Lin},
  \bibinfo{person}{Jheng-Hong Yang}, \bibinfo{person}{Rodrigo Nogueira},
  \bibinfo{person}{Ming-Feng Tsai}, \bibinfo{person}{Chuan-Ju Wang}, {and}
  \bibinfo{person}{Jimmy Lin}.} \bibinfo{year}{2021}\natexlab{}.
\newblock \showarticletitle{Multi-Stage Conversational Passage Retrieval: An
  Approach to Fusing Term Importance Estimation and Neural Query Rewriting}.
\newblock \bibinfo{journal}{\emph{ACM Trans. Inf. Syst.}} \bibinfo{volume}{39},
  \bibinfo{number}{4}, Article \bibinfo{articleno}{48} (\bibinfo{date}{aug}
  \bibinfo{year}{2021}), \bibinfo{numpages}{29}~pages.
\newblock
\showISSN{1046-8188}
\urldef\tempurl%
\url{https://doi.org/10.1145/3446426}
\showDOI{\tempurl}


\bibitem[\protect\citeauthoryear{Liu, Chen, Wu, He, and Zhou}{Liu
  et~al\mbox{.}}{2021}]%
        {9413557}
\bibfield{author}{\bibinfo{person}{Hang Liu}, \bibinfo{person}{Meng Chen},
  \bibinfo{person}{Youzheng Wu}, \bibinfo{person}{Xiaodong He}, {and}
  \bibinfo{person}{Bowen Zhou}.} \bibinfo{year}{2021}\natexlab{}.
\newblock \showarticletitle{Conversational Query Rewriting with Self-Supervised
  Learning}. In \bibinfo{booktitle}{\emph{ICASSP 2021 - 2021 IEEE International
  Conference on Acoustics, Speech and Signal Processing (ICASSP)}}.
  \bibinfo{pages}{7628--7632}.
\newblock
\urldef\tempurl%
\url{https://doi.org/10.1109/ICASSP39728.2021.9413557}
\showDOI{\tempurl}


\bibitem[\protect\citeauthoryear{Mele, Muntean, Nardini, Perego, Tonellotto,
  and Frieder}{Mele et~al\mbox{.}}{2021}]%
        {Mele2021AdaptiveUR}
\bibfield{author}{\bibinfo{person}{Ida Mele}, \bibinfo{person}{Cristina~Ioana
  Muntean}, \bibinfo{person}{Franco~Maria Nardini}, \bibinfo{person}{R.
  Perego}, \bibinfo{person}{Nicola Tonellotto}, {and} \bibinfo{person}{Ophir
  Frieder}.} \bibinfo{year}{2021}\natexlab{}.
\newblock \showarticletitle{Adaptive utterance rewriting for conversational
  search}.
\newblock \bibinfo{journal}{\emph{Inf. Process. Manag.}}  \bibinfo{volume}{58}
  (\bibinfo{year}{2021}), \bibinfo{pages}{102682}.
\newblock


\bibitem[\protect\citeauthoryear{Mele, Tonellotto, Frieder, and Perego}{Mele
  et~al\mbox{.}}{2020}]%
        {ipm2020-MTFP}
\bibfield{author}{\bibinfo{person}{Ida Mele}, \bibinfo{person}{Nicola
  Tonellotto}, \bibinfo{person}{Ophir Frieder}, {and} \bibinfo{person}{Raffaele
  Perego}.} \bibinfo{year}{2020}\natexlab{}.
\newblock \showarticletitle{Topical result caching in web search engines}.
\newblock \bibinfo{journal}{\emph{Inf. Proc. \& Man.}} \bibinfo{volume}{57},
  \bibinfo{number}{3} (\bibinfo{year}{2020}).
\newblock


\bibitem[\protect\citeauthoryear{Neglia, Garetto, and Leonardi}{Neglia
  et~al\mbox{.}}{2022}]%
        {SimCache22}
\bibfield{author}{\bibinfo{person}{Giovanni Neglia}, \bibinfo{person}{Michele
  Garetto}, {and} \bibinfo{person}{Emilio Leonardi}.}
  \bibinfo{year}{2022}\natexlab{}.
\newblock \showarticletitle{Similarity Caching: Theory and Algorithms}.
\newblock \bibinfo{journal}{\emph{IEEE/ACM Transactions on Networking}}
  \bibinfo{volume}{30}, \bibinfo{number}{2} (\bibinfo{year}{2022}),
  \bibinfo{pages}{475--486}.
\newblock
\urldef\tempurl%
\url{https://doi.org/10.1109/TNET.2021.3126368}
\showDOI{\tempurl}


\bibitem[\protect\citeauthoryear{Neyshabur and Srebro}{Neyshabur and
  Srebro}{2015}]%
        {10.5555/3045118.3045323}
\bibfield{author}{\bibinfo{person}{Behnam Neyshabur} {and}
  \bibinfo{person}{Nathan Srebro}.} \bibinfo{year}{2015}\natexlab{}.
\newblock \showarticletitle{On Symmetric and Asymmetric LSHs for Inner Product
  Search}. In \bibinfo{booktitle}{\emph{Proc. ICML}}.
  \bibinfo{pages}{1926–1934}.
\newblock


\bibitem[\protect\citeauthoryear{Qu, Yang, Chen, Qiu, Croft, and Iyyer}{Qu
  et~al\mbox{.}}{2020}]%
        {convdr}
\bibfield{author}{\bibinfo{person}{Chen Qu}, \bibinfo{person}{Liu Yang},
  \bibinfo{person}{Cen Chen}, \bibinfo{person}{Minghui Qiu},
  \bibinfo{person}{W.~Bruce Croft}, {and} \bibinfo{person}{Mohit Iyyer}.}
  \bibinfo{year}{2020}\natexlab{}.
\newblock \bibinfo{booktitle}{\emph{Open-Retrieval Conversational Question
  Answering}}.
\newblock \bibinfo{publisher}{Association for Computing Machinery},
  \bibinfo{address}{New York, NY, USA}, \bibinfo{pages}{539–548}.
\newblock
\showISBNx{9781450380164}
\urldef\tempurl%
\url{https://doi.org/10.1145/3397271.3401110}
\showURL{%
\tempurl}


\bibitem[\protect\citeauthoryear{Reimers and Gurevych}{Reimers and
  Gurevych}{2019}]%
        {sbert}
\bibfield{author}{\bibinfo{person}{Nils Reimers} {and} \bibinfo{person}{Iryna
  Gurevych}.} \bibinfo{year}{2019}\natexlab{}.
\newblock \showarticletitle{{Sentence-BERT: Sentence Embeddings using Siamese
  BERT-Networks}}. In \bibinfo{booktitle}{\emph{Proc. EMNLP}}.
  \bibinfo{pages}{3980--3990}.
\newblock


\bibitem[\protect\citeauthoryear{Sermpezis, Giannakas, Spyropoulos, and
  Vigneri}{Sermpezis et~al\mbox{.}}{2018}]%
        {SoftCache}
\bibfield{author}{\bibinfo{person}{Pavlos Sermpezis},
  \bibinfo{person}{Theodoros Giannakas}, \bibinfo{person}{Thrasyvoulos
  Spyropoulos}, {and} \bibinfo{person}{Luigi Vigneri}.}
  \bibinfo{year}{2018}\natexlab{}.
\newblock \showarticletitle{Soft Cache Hits: Improving Performance Through
  Recommendation and Delivery of Related Content}.
\newblock \bibinfo{journal}{\emph{IEEE Journal on Selected Areas in
  Communications}} \bibinfo{volume}{36}, \bibinfo{number}{6}
  (\bibinfo{year}{2018}), \bibinfo{pages}{1300--1313}.
\newblock
\urldef\tempurl%
\url{https://doi.org/10.1109/JSAC.2018.2844983}
\showDOI{\tempurl}


\bibitem[\protect\citeauthoryear{Vakulenko, Longpre, Tu, and Anantha}{Vakulenko
  et~al\mbox{.}}{2021}]%
        {10.1145/3437963.3441748}
\bibfield{author}{\bibinfo{person}{Svitlana Vakulenko}, \bibinfo{person}{Shayne
  Longpre}, \bibinfo{person}{Zhucheng Tu}, {and} \bibinfo{person}{Raviteja
  Anantha}.} \bibinfo{year}{2021}\natexlab{}.
\newblock \showarticletitle{Question Rewriting for Conversational Question
  Answering}. In \bibinfo{booktitle}{\emph{Proceedings of the 14th ACM
  International Conference on Web Search and Data Mining}}
  \emph{(\bibinfo{series}{WSDM '21})}. \bibinfo{publisher}{Association for
  Computing Machinery}, \bibinfo{address}{New York, NY, USA},
  \bibinfo{pages}{355–363}.
\newblock
\showISBNx{9781450382977}
\urldef\tempurl%
\url{https://doi.org/10.1145/3437963.3441748}
\showDOI{\tempurl}


\bibitem[\protect\citeauthoryear{van~der Maaten and Hinton}{van~der Maaten and
  Hinton}{2008}]%
        {TSNE}
\bibfield{author}{\bibinfo{person}{Laurens van~der Maaten} {and}
  \bibinfo{person}{Geoffrey Hinton}.} \bibinfo{year}{2008}\natexlab{}.
\newblock \showarticletitle{Visualizing Data using t-SNE}.
\newblock \bibinfo{journal}{\emph{Journal of Machine Learning Research}}
  \bibinfo{volume}{9}, \bibinfo{number}{86} (\bibinfo{year}{2008}),
  \bibinfo{pages}{2579--2605}.
\newblock


\bibitem[\protect\citeauthoryear{Voskarides, Li, Ren, Kanoulas, and
  de~Rijke}{Voskarides et~al\mbox{.}}{2020}]%
        {10.1145/3397271.3401130}
\bibfield{author}{\bibinfo{person}{Nikos Voskarides}, \bibinfo{person}{Dan Li},
  \bibinfo{person}{Pengjie Ren}, \bibinfo{person}{Evangelos Kanoulas}, {and}
  \bibinfo{person}{Maarten de Rijke}.} \bibinfo{year}{2020}\natexlab{}.
\newblock \bibinfo{booktitle}{\emph{Query Resolution for Conversational Search
  with Limited Supervision}}.
\newblock \bibinfo{publisher}{Association for Computing Machinery},
  \bibinfo{address}{New York, NY, USA}, \bibinfo{pages}{921–930}.
\newblock
\showISBNx{9781450380164}
\urldef\tempurl%
\url{https://doi.org/10.1145/3397271.3401130}
\showURL{%
\tempurl}


\bibitem[\protect\citeauthoryear{Xiong, Xiong, Li, Tang, Liu, Bennett, Ahmed,
  and Overwijk}{Xiong et~al\mbox{.}}{2021}]%
        {ance}
\bibfield{author}{\bibinfo{person}{Lee Xiong}, \bibinfo{person}{Chenyan Xiong},
  \bibinfo{person}{Ye Li}, \bibinfo{person}{Kwok-Fung Tang},
  \bibinfo{person}{Jialin Liu}, \bibinfo{person}{Paul Bennett},
  \bibinfo{person}{Junaid Ahmed}, {and} \bibinfo{person}{Arnold Overwijk}.}
  \bibinfo{year}{2021}\natexlab{}.
\newblock \showarticletitle{Approximate Nearest Neighbor Negative Contrastive
  Learning for Dense Text Retrieval}. In \bibinfo{booktitle}{\emph{Proc.
  ICLR}}.
\newblock


\bibitem[\protect\citeauthoryear{Xu, Moon, Liu, Liu, Shah, Liu, and Yu}{Xu
  et~al\mbox{.}}{2020}]%
        {xu-etal-2020-user}
\bibfield{author}{\bibinfo{person}{Hu Xu}, \bibinfo{person}{Seungwhan Moon},
  \bibinfo{person}{Honglei Liu}, \bibinfo{person}{Bing Liu},
  \bibinfo{person}{Pararth Shah}, \bibinfo{person}{Bing Liu}, {and}
  \bibinfo{person}{Philip Yu}.} \bibinfo{year}{2020}\natexlab{}.
\newblock \showarticletitle{User Memory Reasoning for Conversational
  Recommendation}. In \bibinfo{booktitle}{\emph{Proceedings of the 28th
  International Conference on Computational Linguistics}}.
  \bibinfo{publisher}{International Committee on Computational Linguistics},
  \bibinfo{address}{Barcelona, Spain (Online)}, \bibinfo{pages}{5288--5308}.
\newblock
\urldef\tempurl%
\url{https://doi.org/10.18653/v1/2020.coling-main.463}
\showDOI{\tempurl}


\bibitem[\protect\citeauthoryear{Yang, Lin, Wang, Lin, and Tsai}{Yang
  et~al\mbox{.}}{2019}]%
        {Yang2019QueryAA}
\bibfield{author}{\bibinfo{person}{Jheng-Hong Yang},
  \bibinfo{person}{Sheng-Chieh Lin}, \bibinfo{person}{Chuan-Ju Wang},
  \bibinfo{person}{Jimmy~J. Lin}, {and} \bibinfo{person}{Ming-Feng Tsai}.}
  \bibinfo{year}{2019}\natexlab{}.
\newblock \showarticletitle{Query and Answer Expansion from Conversation
  History}. In \bibinfo{booktitle}{\emph{TREC}}.
\newblock


\bibitem[\protect\citeauthoryear{Yu, Liu, Yang, Xiong, Bennett, Gao, and
  Liu}{Yu et~al\mbox{.}}{2020}]%
        {10.1145/3397271.3401323}
\bibfield{author}{\bibinfo{person}{Shi Yu}, \bibinfo{person}{Jiahua Liu},
  \bibinfo{person}{Jingqin Yang}, \bibinfo{person}{Chenyan Xiong},
  \bibinfo{person}{Paul Bennett}, \bibinfo{person}{Jianfeng Gao}, {and}
  \bibinfo{person}{Zhiyuan Liu}.} \bibinfo{year}{2020}\natexlab{}.
\newblock \bibinfo{booktitle}{\emph{Few-Shot Generative Conversational Query
  Rewriting}}.
\newblock \bibinfo{publisher}{Association for Computing Machinery},
  \bibinfo{address}{New York, NY, USA}, \bibinfo{pages}{1933–1936}.
\newblock
\showISBNx{9781450380164}
\urldef\tempurl%
\url{https://doi.org/10.1145/3397271.3401323}
\showURL{%
\tempurl}


\bibitem[\protect\citeauthoryear{Yu, Liu, Xiong, Feng, and Liu}{Yu
  et~al\mbox{.}}{2021}]%
        {CastDR}
\bibfield{author}{\bibinfo{person}{Shi Yu}, \bibinfo{person}{Zhenghao Liu},
  \bibinfo{person}{Chenyan Xiong}, \bibinfo{person}{Tao Feng}, {and}
  \bibinfo{person}{Zhiyuan Liu}.} \bibinfo{year}{2021}\natexlab{}.
\newblock \showarticletitle{Few-Shot Conversational Dense Retrieval}. In
  \bibinfo{booktitle}{\emph{Proc. SIGIR}}. \bibinfo{pages}{829–838}.
\newblock


\bibitem[\protect\citeauthoryear{Zamani, Trippas, Dalton, and Radlinski}{Zamani
  et~al\mbox{.}}{2022}]%
        {zamani2022conversational}
\bibfield{author}{\bibinfo{person}{Hamed Zamani}, \bibinfo{person}{Johanne~R
  Trippas}, \bibinfo{person}{Jeff Dalton}, {and} \bibinfo{person}{Filip
  Radlinski}.} \bibinfo{year}{2022}\natexlab{}.
\newblock \showarticletitle{Conversational Information Seeking}.
\newblock \bibinfo{journal}{\emph{arXiv:2201.08808}} (\bibinfo{year}{2022}).
\newblock


\bibitem[\protect\citeauthoryear{Zhan, Mao, Liu, Guo, Zhang, and Ma}{Zhan
  et~al\mbox{.}}{2021}]%
        {star}
\bibfield{author}{\bibinfo{person}{Jingtao Zhan}, \bibinfo{person}{Jiaxin Mao},
  \bibinfo{person}{Yiqun Liu}, \bibinfo{person}{Jiafeng Guo},
  \bibinfo{person}{Min Zhang}, {and} \bibinfo{person}{Shaoping Ma}.}
  \bibinfo{year}{2021}\natexlab{}.
\newblock \showarticletitle{Optimizing Dense Retrieval Model Training with Hard
  Negatives}. In \bibinfo{booktitle}{\emph{Proc. SIGIR}}.
  \bibinfo{pages}{1503–1512}.
\newblock


\bibitem[\protect\citeauthoryear{Zhang, Chen, Ai, Yang, and Croft}{Zhang
  et~al\mbox{.}}{2018}]%
        {10.1145/3269206.3271776}
\bibfield{author}{\bibinfo{person}{Yongfeng Zhang}, \bibinfo{person}{Xu Chen},
  \bibinfo{person}{Qingyao Ai}, \bibinfo{person}{Liu Yang}, {and}
  \bibinfo{person}{W.~Bruce Croft}.} \bibinfo{year}{2018}\natexlab{}.
\newblock \showarticletitle{Towards Conversational Search and Recommendation:
  System Ask, User Respond}. In \bibinfo{booktitle}{\emph{Proceedings of the
  27th ACM International Conference on Information and Knowledge Management}}
  \emph{(\bibinfo{series}{CIKM '18})}. \bibinfo{publisher}{Association for
  Computing Machinery}, \bibinfo{address}{New York, NY, USA},
  \bibinfo{pages}{177–186}.
\newblock
\showISBNx{9781450360142}
\urldef\tempurl%
\url{https://doi.org/10.1145/3269206.3271776}
\showDOI{\tempurl}


\end{thebibliography}


\end{document}